\documentclass[11pt,a4paper]{scrartcl}

\usepackage{CLICdp}

\usepackage{CLICdp_definitions}
\usepackage{multirow}

\usepackage[percent]{overpic}
\newcommand{\ddhep}{\textsc{DD4hep}\xspace}
\newcommand{\ddg}{\textsc{DDG4}\xspace}

\addto\extrasenglish{%
}

\title{Optimization of timing selections at 380~GeV CLIC}

\clicdpnote{2018}{003}

\date{\today}

\addauthor{E.~Brondolin\thanks{erica.brondolin@cern.ch}}{\institute{1}}
\addauthor{A.~Sailer\thanks{andre.philippe.sailer@cern.ch}}{\institute{1}}

\addinstitute{1}{CERN, Switzerland}

\abstract{The Compact Linear Collider (CLIC) is a proposed high-luminosity linear electron-positron collider at the energy frontier. 
It is foreseen to be built and operated in three stages, with a centre-of-mass energy ranging from a few hundred GeV up to 3 TeV.
The main beam-induced background impacting CLIC physics analyses is produced by beamstrahlung radiation from the electron and positron bunches
traversing the high field of the opposite beam and converting to hadrons, \gghad{} events.
The timing selections are a powerful tool to discriminate signal from background events at CLIC.
For each CLIC stage, three sets of selections, \textit{Loose}, \textit{Selected}, and \textit{Tight}, 
are defined to allow the analysis of different signal topologies.
The selections are defined depending on the particle type and the reconstructed polar angle 
of each particle flow object reconstructed using the particle flow analysis. 
As a first step, the performance of the timing selections currently defined for the CLIC \SI{380}{\GeV} energy stage are evaluated
using a \ttbar sample decaying mainly into light quarks as signal and \gghad{} events as background.
As a result, after applying the selections the level of background is significantly reduced, down to few~GeV, 
while the shift of the signal peak to lower energy is kept within a few percent.
The cuts in the \textit{Loose} selection are then relaxed to search for possible improvements:
in one of the options considered, the signal component is partially recovered
but also the background energy mean remains unchanged with respect to not applying any cuts.
In conclusion, no change is foreseen on the timing selections.
}

\titlecomment{This work was carried out in the framework of the CLICdp Collaboration} 

\graphicspath{ {./logos/}{./figures/} }

\begin{document}

\titlepage

\section{Introduction}\label{sec:Intro}

The Compact Linear Collider (CLIC) is a proposed high-luminosity linear electron--positron collider at the energy frontier~\cite{cdrvol2, cdrupdate}.
It is foreseen to be built and operated in three stages with increasing centre-of-mass energies.
In the first stage, CLIC will mainly operate at a centre-of-mass energy of \SI{380}{\GeV} with the 
aim of measuring the properties of the top quark and the Higgs boson with high precision.
The accelerator parameters for the first stage are reported in~\autoref{tab:accelerator_parameters}.
The background at CLIC is produced by beamstrahlung radiation from the electron and positron bunches
traversing the high field of the opposite beam. Photons then convert to two main types of background, incoherent \epem pairs and \gghad{} events. 
At the \SI{380}{\GeV} energy stage, 0.18 \gghad{} interactions occur on average per bunch crossing with centre-of-mass energy more than \SI{2}{\GeV}~\cite{clic-beam-beam}.

\begin{figure}
  \centering
  \begin{overpic}[width=.8\textwidth]{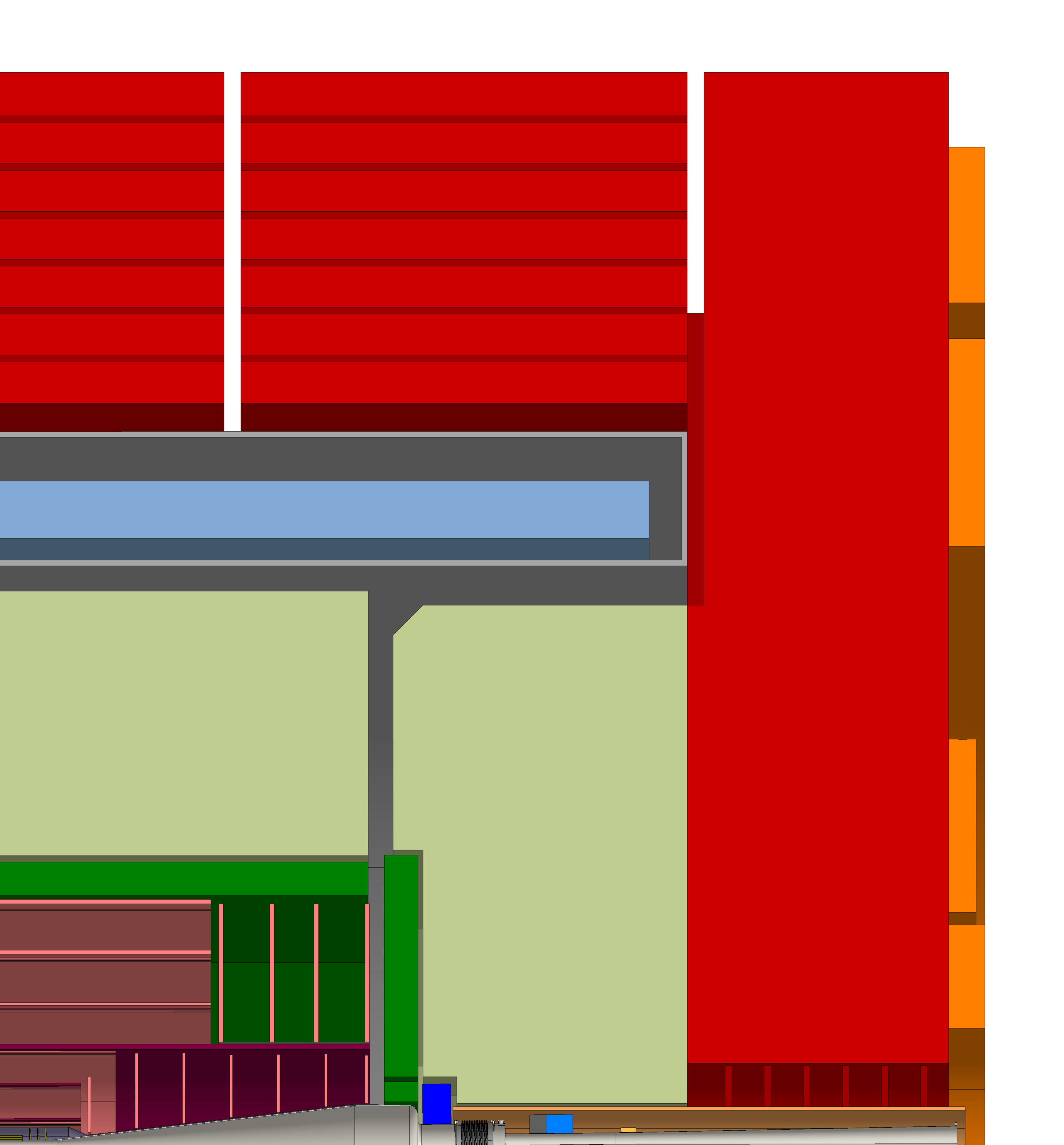}
  \put (30,77.5) {\huge{Yoke + muon chambers}}
  \put (20,54) {\huge{Solenoid}}
  \put (10,35) {\huge{HCAL}}
  \put (12,22.5) {\large{ECAL}}
  \put (5,13.5) {\color{white}\large{Tracker}}
  \put (0,0) {\color{white}\small{Vertex}}
  \end{overpic}
  \caption{Longitudinal cross section of CLICdet~\cite{CLICperf}.}
  \label{fig:CLICdet_quarter}
\end{figure}
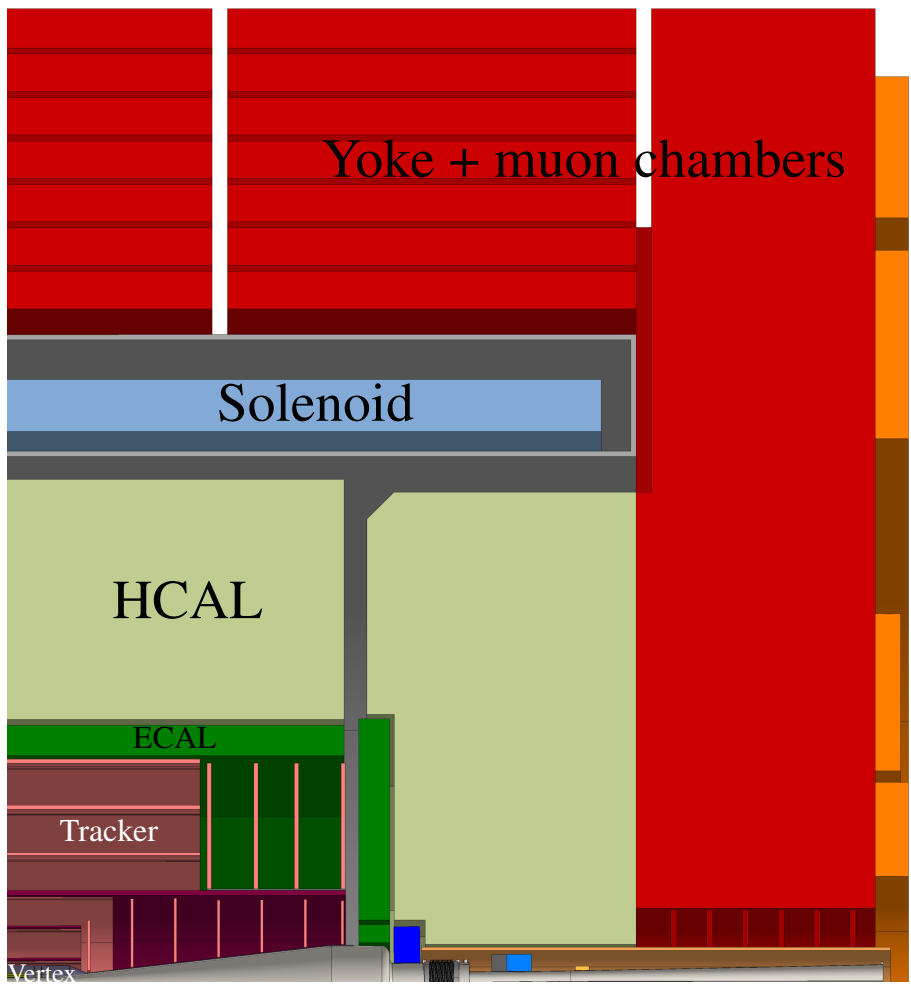

To face the challenge of the beam-induced background and to fulfil the physics requirements, a detector model 
with a highly segmented calorimetry system optimized for particle flow techniques and timing capabilities has been defined
and optimized using a dedicated software suite~\cite{CLICdet}. 
The CLICdet layout follows the typical collider detector scheme: 
the innermost part is composed of a silicon pixel vertex detector and a silicon tracker;
surrounding them, an electromagnetic and hadronic calorimeter
are placed, all embedded inside a superconducting solenoid providing a \SI{4}{\tesla} field; 
in the outermost part an iron yoke is interleaved with muon chambers.
A longitudinal cross section of CLICdet is shown in~\autoref{fig:CLICdet_quarter}.

CLICdet will operate in a triggerless readout mode, i.e. 
the entire bunch train composed of 352 bunches at the \SI{380}{\GeV} energy stage
will be read out every \SI{20}{\ms}. In the bunch train, the bunches are separated by \SI{0.5}{\ns}.
At most, one hard \epem physics interaction in an entire bunch train will be produced,
while most of the bunch crossings will only produce background particles.
To obtain excellent detector performance in this triggerless readout mode, 
the subdetectors must provide a precise hit timing information.
This can be achieved with a time-stamping capability of
\SI{10}{\ns} for all silicon tracking elements and a hit time resolution of \SI{1}{\ns} for all calorimeter hits.
Studies have shown~\cite{cdrvol2} that at the highest energy stage of CLIC,
the timing information coming from the subdetectors combined with additional \pT{} information
can help to mitigate the impact of the \gghads{} background.
In the context of a \SI{380}{\GeV} CLIC collider, a preliminary set of timing selections are already defined
but the performance is studied precisely for the first time in this note.
Therefore, as a first step the performance of the timing selections currently used for the first CLIC stage 
are evaluated in terms of absolute acceptance of the signal and rejection of the background;
while in a second step, the loosest selection available is relaxed to check if further optimization is possible.
Since a precise measurement of the top quark is one of the most important parts in the CLIC physics programme
at the first stage, a pair of top quarks produced at the \epem interaction point is used as the signal.

The software framework and event simulation used in this study is briefly described in~\cref{sec:simreco}.
In~\cref{sec:timingCuts} an introduction to the timing selections is given 
with particular emphasis on those used for the first CLIC stage.
Further possible optimizations for the loose selection are described in~\cref{sec:LEtimingCuts}.
In~\cref{sec:results} this study is summarized and conclusions are given.

\begin{table}
\centering
\caption{Parameters for the first stage of CLIC.}
\label{tab:accelerator_parameters}
\begin{tabular}{l c }
\toprule
Parameter  & Stage 1 \\
\midrule
Centre-of-mass energy [\si{\GeV}] & 380 \\
Main tunnel length [\si{\km}] & 11.4 \\
Repetition frequency [\si{\Hz}] & 50 \\
Number of bunches & 352 \\
Bunch separation [\si{\ns}] & 0.5 \\
Number of particles per bunch [\num{e9}] & 5.2 \\
IP beam size (horizontal) [\si{\nm}] & $\simeq$149 \\
IP beam size (vertical) [\si{\nm}] & 2.9 \\
Bunch length [\si{\micron}] & 70 \\
Total luminosity [\SI{e34}{\per\square\cm\per\s}] & 1.5 \\
Luminosity above 99\% of $\sqrt{s}$ [\SI{e34}{\per\square\cm\per\s}] & 0.9 \\
Number of beamstrahlung photons per beam particle & 1.4 \\
Number of \gghad{} events per BX ($\sqrt{s} > \SI{2}{\GeV}$) & 0.18 \\
\bottomrule
\end{tabular}
\end{table}

\section{Simulation and reconstruction}
\label{sec:simreco}

The CLICdet detector geometry is described with the \ddhep{} software framework~\cite{frank13:dd4hep} and simulated in \geant{}~\cite{AGOSTINELLI2003250} 
via the \ddg{}~\cite{frank15:ddg4} package of \ddhep{}.

The simulated events are generated mostly using the \whizard{}~\cite{Kilian:2007gr,Moretti:2001zz} 
program assuming zero polarization of the electron and proton beam.
The initial states are created in the \guineapig{}~\cite{Schulte:1997nga} simulation of the CLIC collisions.
The performance of the timing selections for the \SI{380}{\GeV} CLIC stage is assessed 
using a 6-fermion production, comprising predominantly \ttbar events from hard Standard Model \epem interactions. 
The six fermions in the final state correspond to $\PQd\PQd\PQu yy \PQu$, where $y$ can be b, d or s quarks.
The simulation of the parton showering, hadronisation and fragmentation is performed using \pythia{}~\cite{Sjostrand:2006za}.
This component of the event is called \textit{signal}. 
The \gghad{} events are hadronized in \pythia{} and the amount correspondent to 30 bunch crossings (BX) is included.
This component of the event is referred to as \textit{background}.

The reconstruction software is implemented in the linear collider \marlin{}-framework~\cite{MarlinLCCD}.
The first step of the reconstruction is to overlay the signal event with the expected number of \gghad{} background events
for 30 bunch crossings. The signal event is placed in bunch crossing 11 at $t=0~\si{\ns}$. 
At this stage, only the energy deposits inside the timing window of \SI{10}{\ns} after the signal event are selected.
This time window matches the expected detector timing resolutions and integration times.
In the next step, the positions of the energy deposits in the tracking detectors are smeared and the energy of the
calorimeter hits are scaled with the calibration constants.
At this point the tracking algorithms are run to obtain reconstructed tracks
and the particle flow approach by \pandora~\cite{Marshall:2015rfaPandoraSDK,Thomson:2009rp,Marshall:2012ryPandoraPFA}
is used to build calorimeter clusters and match them to tracks.
All visible particles are reconstructed by \pandora using both reconstructed tracks and calorimeter hits.
The particles are called \textit{particle flow objects} (PFOs).
A PFO is marked as signal if any hit in its cluster are coming from a MC Particle of the \ttbar event.
If this requirement is not satisfied, it is marked as background.
A more detailed description of the reconstruction of single particles and more complex events is given in~\cite{CLICperf}.

\section{Timing selections}
\label{sec:timingCuts}

\begin{figure}
  \centering
  \includegraphics[width=.48\textwidth]{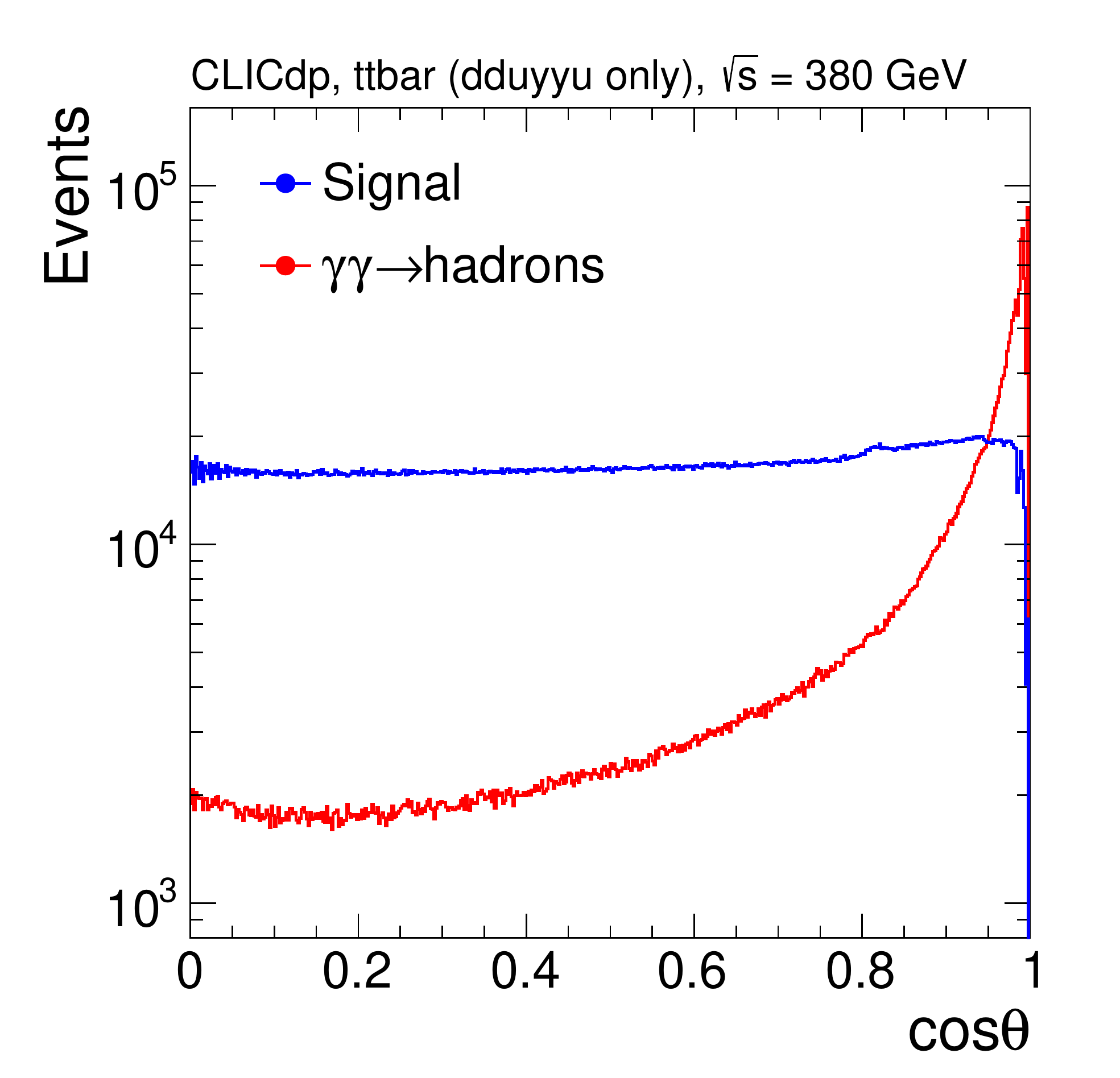}
  \caption{Reconstructed polar angle of all PFO reconstructed in the event, marked as
  signal (blue) and the \gghads{} (red).}
  \label{fig:costheta}
\end{figure}

To reject the beam-induced background without impacting the physics performance,
the time-stamping capabilities of CLICdet described in~\cref{sec:Intro} are used.
Simply imposing tight timing selections at the hit level does not provide a final solution
due to the fact that it does not take into account the hadronic shower development time and 
the time-of-flight corrections for lower momentum particles.
Only the combination of both timing capabilities of the high-granularity calorimeters and particle flow reconstruction
allows one to obtain a precise time-stamp for the calorimeter clusters.
Once the particle flow clustering is finished, \pT{} dependent timing selections are applied.
The timing selections are therefore also called \pT{} vs. time selections and their main
aim is to reduce the background at the minimum possible while keeping invariant the signal.
They are imposed to each reconstructed PFO and are defined depending on the particle type
and its reconstructed polar angle. The reconstructed polar angle is shown in~\cref{fig:costheta}
for both the signal and the \gghads{}.

PFOs are divided into three exclusive categories -- photons, neutral hadrons and charged particles --
and accepted or rejected based on the time of the calorimeter clusters and their reconstructed transverse momentum.
The cluster time is computed using the truncated energy weighted mean time of its calorimeter hits.
By default, the cluster time is computed using HCAL clusters. ECAL clusters are used only in the following cases:
\begin{itemize}
\item if the PFO is a photon;
\item if the number of ECAL hits reconstructed for the PFO is more than five;
\item if the number of ECAL hits reconstructed for the PFO is more or equal than half of the number of HCAL hits.
\end{itemize}
The cluster time is corrected using the time-of-flight information coming from the associated track
in the case of a charged particle, and the time-of-flight of a straight path in the case of neutrals.

\begin{figure}
  \centering
  \begin{subfigure}[b]{0.48\textwidth}
    \includegraphics[width=\textwidth]{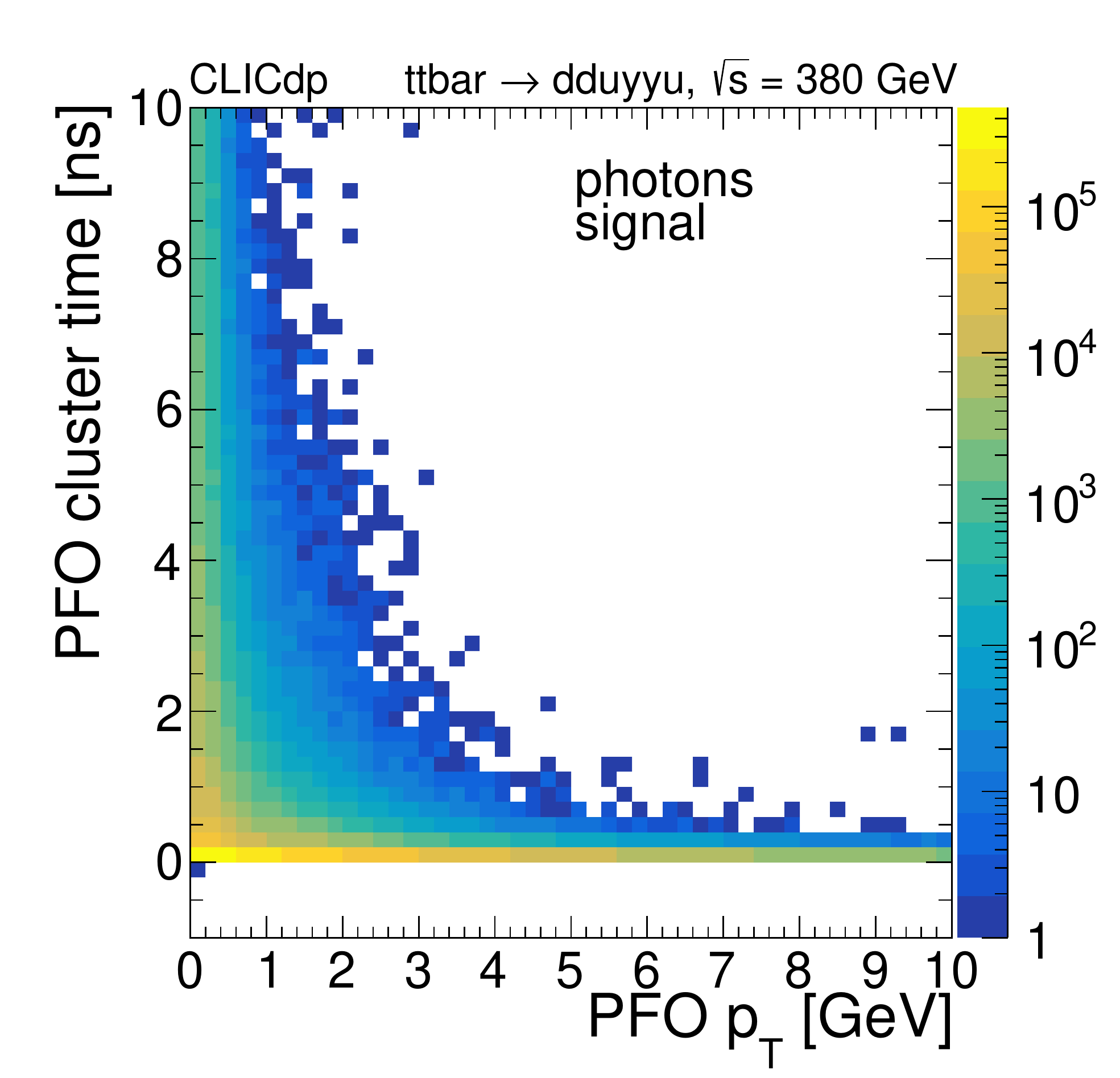}
  \end{subfigure}
  \hfill
  \begin{subfigure}[b]{0.48\textwidth}
    \includegraphics[width=\textwidth]{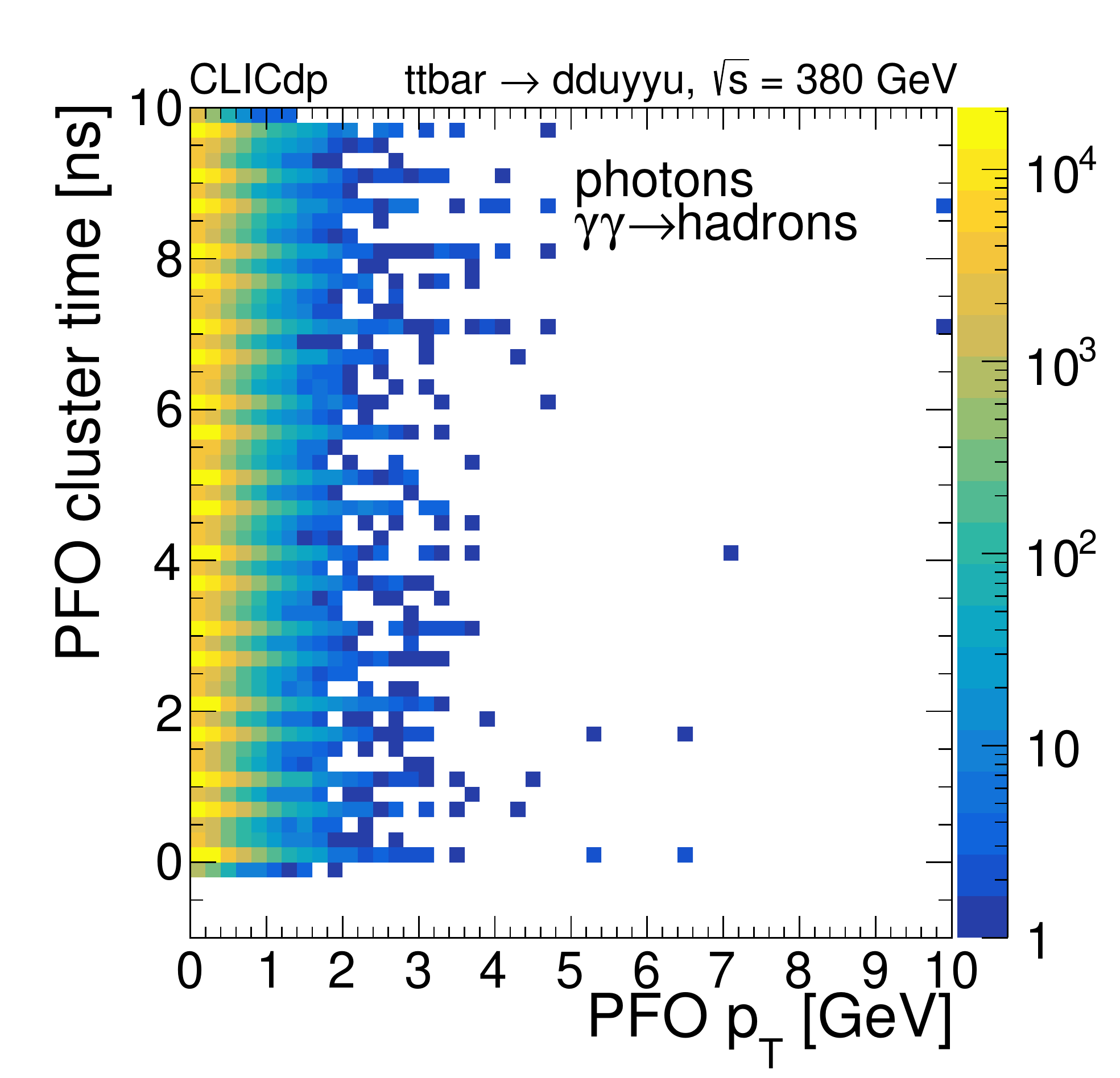}
  \end{subfigure}
  \begin{subfigure}[b]{0.48\textwidth}
    \includegraphics[width=\textwidth]{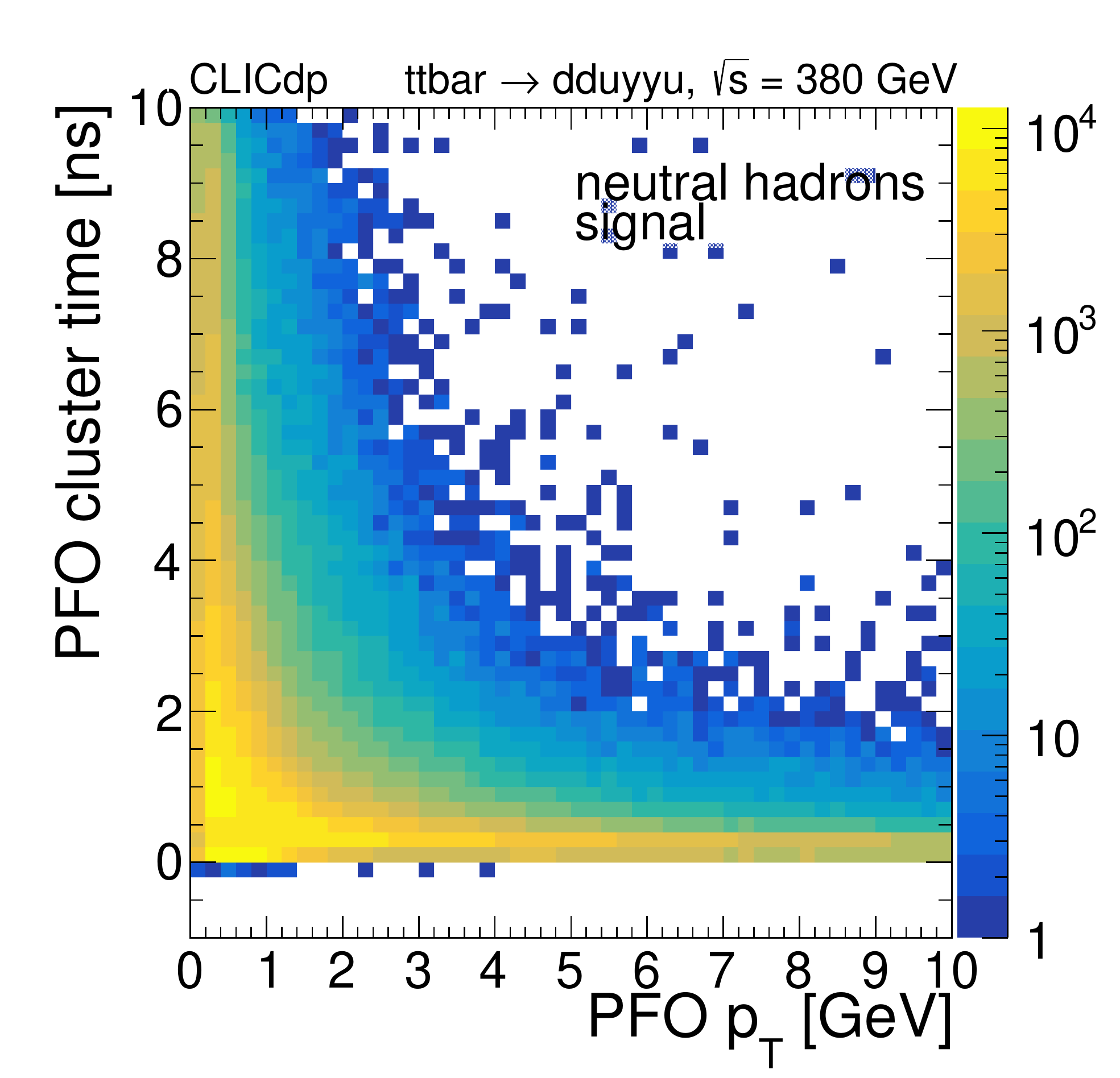}
  \end{subfigure}
  \hfill
  \begin{subfigure}[b]{0.48\textwidth}
    \includegraphics[width=\textwidth]{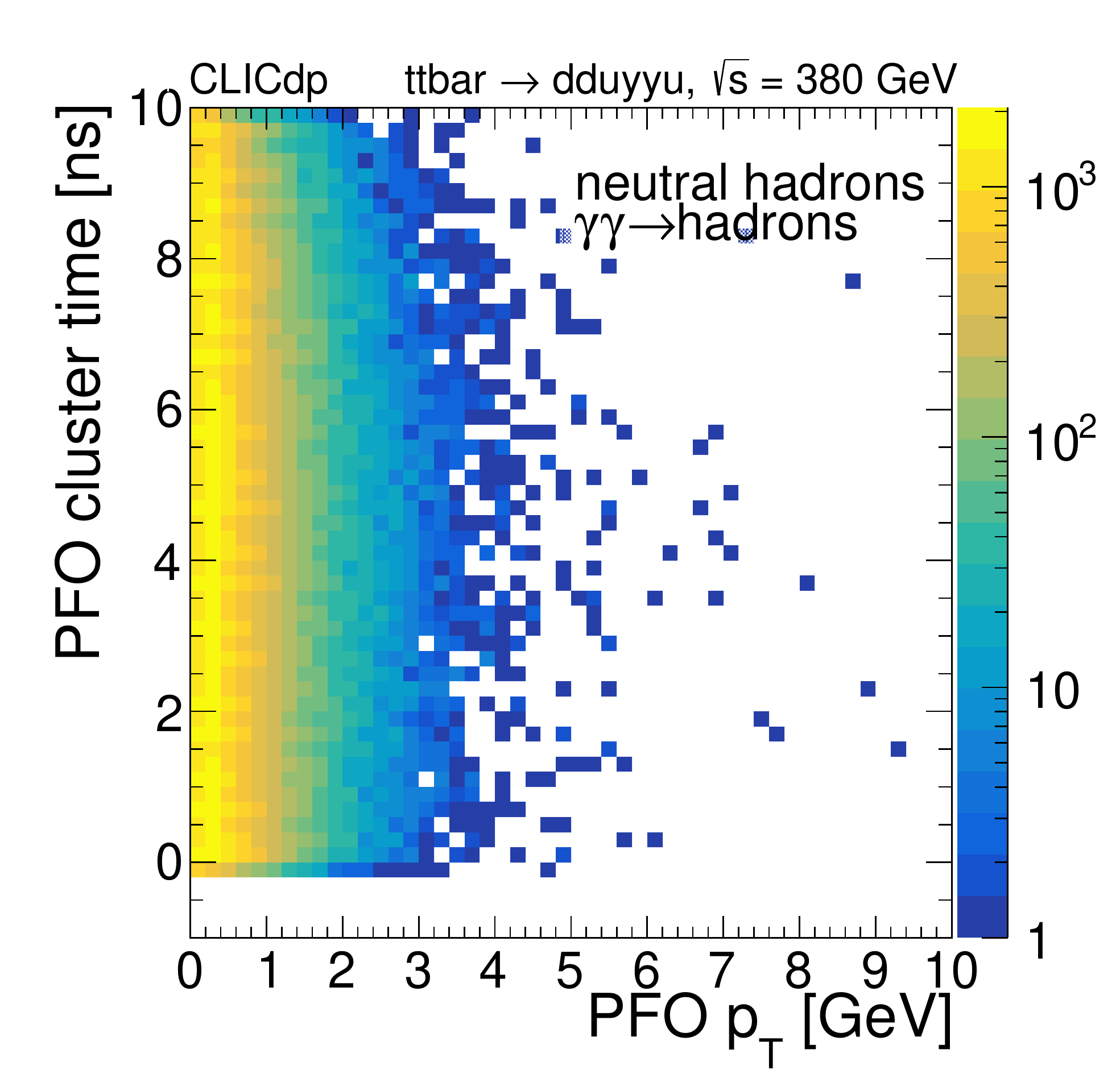}
  \end{subfigure}
  \begin{subfigure}[b]{0.48\textwidth}
    \includegraphics[width=\textwidth]{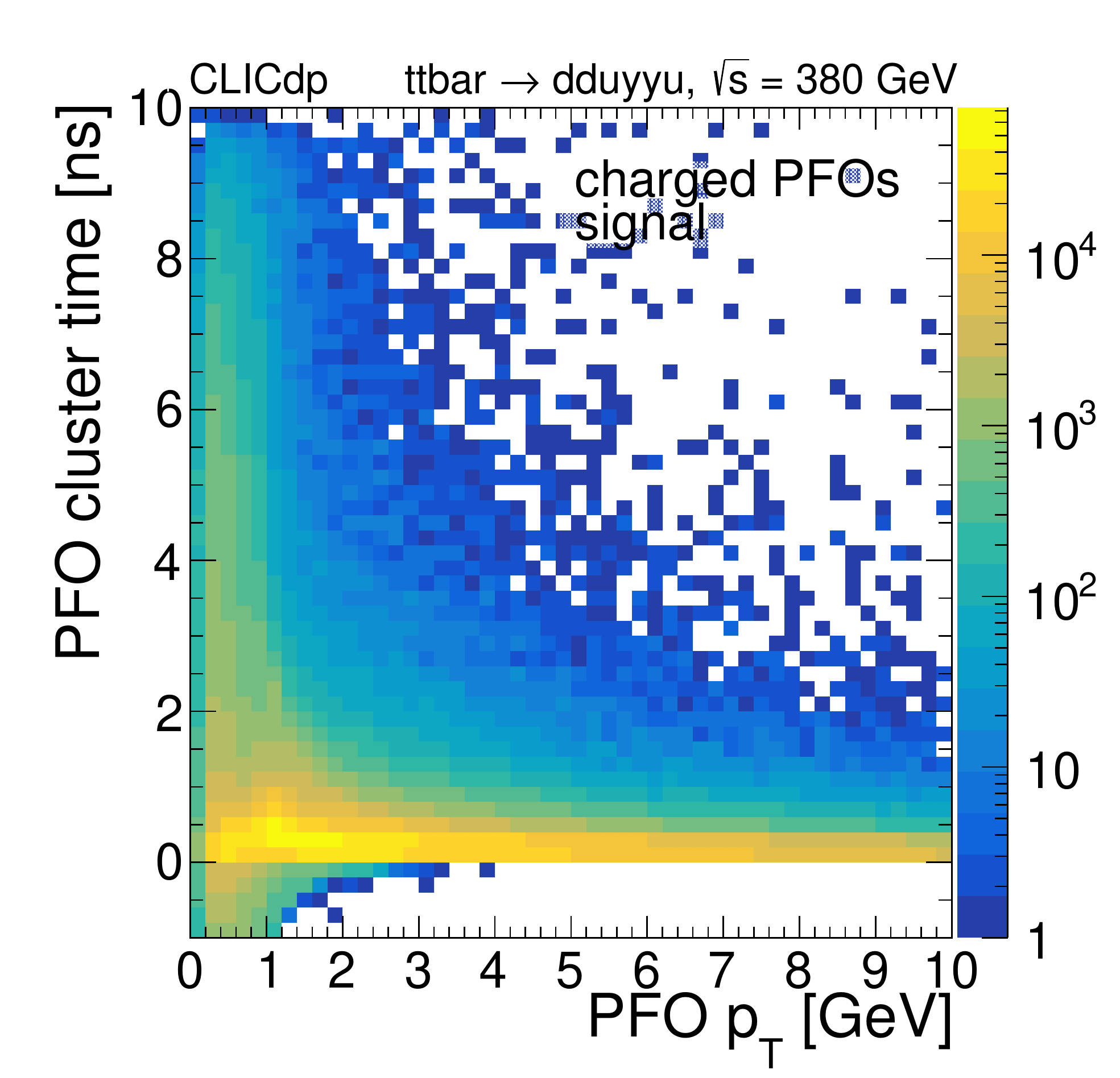}
  \end{subfigure}
  \hfill
  \begin{subfigure}[b]{0.48\textwidth}
    \includegraphics[width=\textwidth]{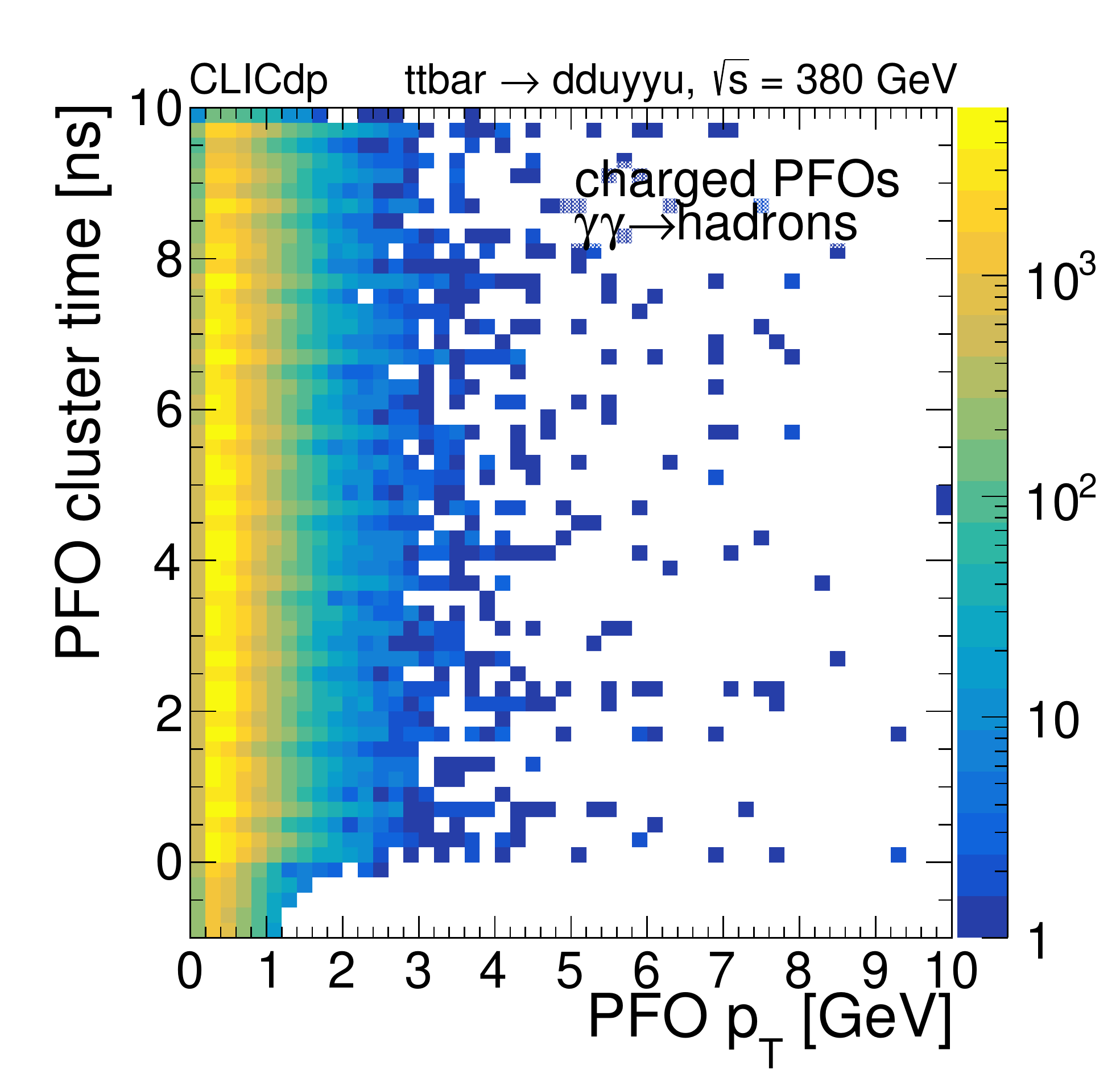}
  \end{subfigure}
  \caption{Cluster time as a function of the \pT{} of the three particle categories (photons at the top,
  neutral hadrons in the middle and charged PFOs at the bottom) for both the signal (left)
  and the background (right) components of \ttbar events with overlay of 30 BX of \gghads{} background for the \SI{380}{\GeV} CLIC stage.}
  \label{fig:timepT_ttbar_380GeV}
\end{figure}

\begin{figure}
  \centering
  \includegraphics[width=\textwidth]{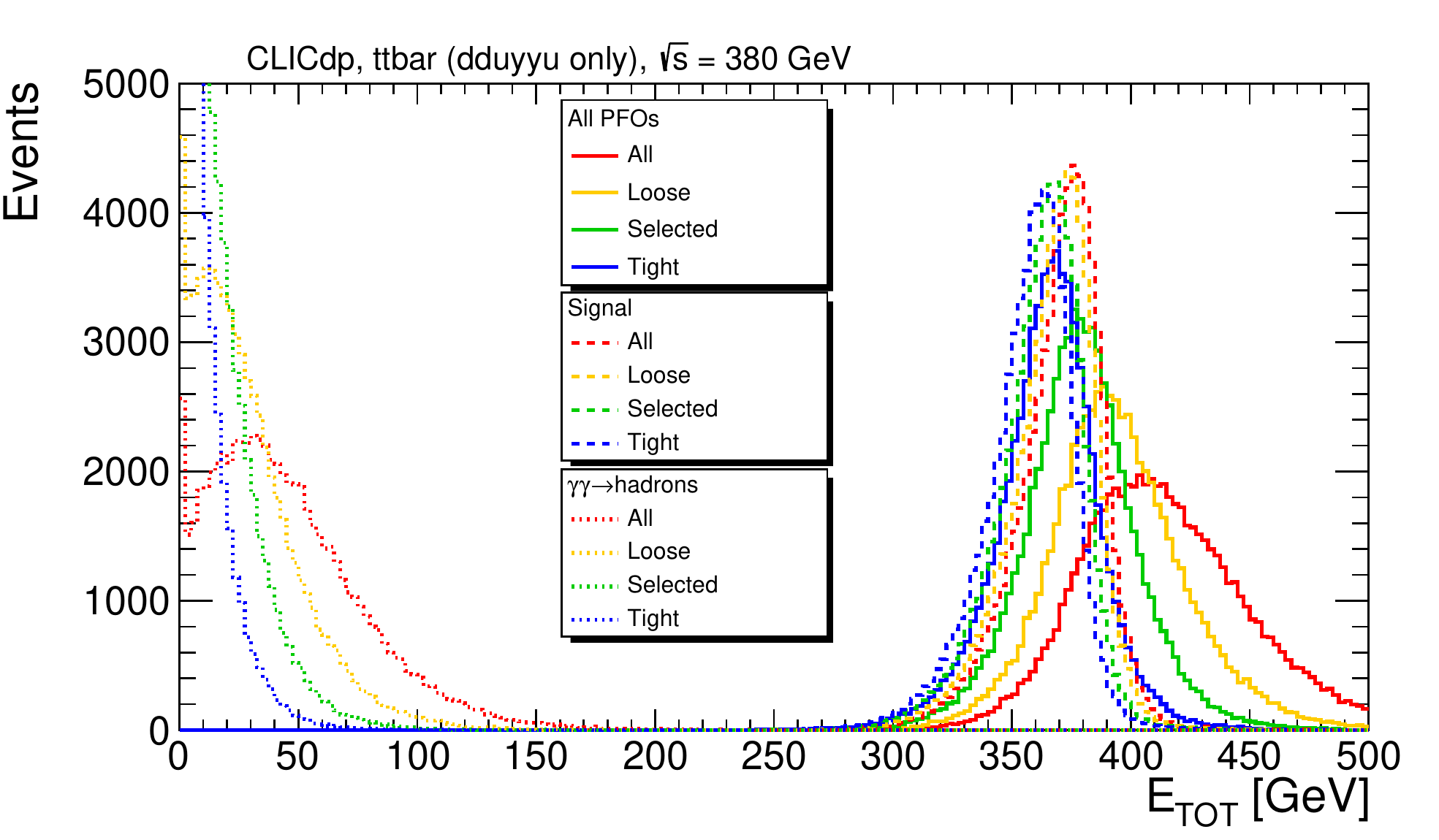}
  \caption{Reconstructed energy distribution for all PFOs (solid line), and for PFOs passing the different timing cuts for signal (dashed line)
  and background (dotted line): without any selection (red), \textit{Loose} (orange), \textit{Selected} (green) and \textit{Tight} (blue). 
  \ttbar events are used with overlay of 30 BX of \gghads{} background for the \SI{380}{\GeV} CLIC stage.}
  \label{fig:energy_ttbar_380GeV_cuts}
\end{figure}

In~\autoref{fig:timepT_ttbar_380GeV} the cluster time as a function of the \pT{} of the three particle categories for both the signal 
and the background components of \ttbar events is shown for the \SI{380}{\GeV} CLIC collider.
Given the large overlap of the signal and background distributions, it is clear that the search for highly performant selections is a
challenging task. Three sets of \pT{} vs. time selections are therefore created to allow the analysis of different signal topologies:
\textit{Loose}, \textit{Selected}, and \textit{Tight}. In all selections, no photon or neutral hadron with \pT{} more than \SI{2}{\GeV}
is rejected, and no timing cut is applied on charged particles if their \pT{} is more than \SI{4}{\GeV} in the case of
\textit{Loose} and \textit{Selected} selections, and more than \SI{3}{\GeV} in the case of \textit{Tight} one.
The three selections for the \SI{380}{\GeV} CLIC stage are listed in~\cref{tab:LE_loose,tab:LE_sel,tab:LE_tight}.
The effect of the three timing selections on the reconstructed energy for both the signal 
and the background components of \ttbar events with overlay of 30 BX of \gghads{} background is shown in~\autoref{fig:energy_ttbar_380GeV_cuts}.

\begin{table}
\centering
\caption{\textit{Loose} selections used for the \SI{380}{\GeV} CLIC stage.}
\label{tab:LE_loose}

\begin{tabular}{lccccc}
  \toprule
  & 
  \multicolumn{2}{c}{$|\cos\theta| \leq 0.975$} & &
  \multicolumn{2}{c}{$|\cos\theta| > 0.975$} \\ \cmidrule{2-3} \cmidrule{5-6}
  &
  \pT{} range [\si{\GeV}]& time cut [\si{\ns}] & &
  \pT{} range [\si{\GeV}]& time cut [\si{\ns}] \\
  \cmidrule{2-3}  \cmidrule{5-6}
  \multirow{2}{*}{Photons} &
  $0.75 \leq \pT{} < 2.00 $ & $t < 10.0$ &  &
  $0.75 \leq \pT{} < 2.00 $ & $t < 2.00$ \\
   & 
  $0.00 \leq \pT{} < 0.75 $ & $t < 2.50$ & &
  $0.00 \leq \pT{} < 0.75 $ & $t < 1.00$ \\
  \cmidrule{2-3}  \cmidrule{5-6}
  \multirow{2}{*}{Neutral hadrons} &
  $0.75 \leq \pT{} < 2.00 $ & $t < 10.0$ &  &
  $0.75 \leq \pT{} < 2.00 $ & $t < 10.0$ \\
   & 
  $0.00 \leq \pT{} < 0.75 $ & $t < 5.00$ & &
  $0.00 \leq \pT{} < 0.75 $ & $t < 5.00$ \\
  \cmidrule{2-6}
  & 
  \multicolumn{5}{c}{$0 \leq |\cos\theta| \leq 1$} \\  \cmidrule{3-5}
  & &  \pT{} range [\si{\GeV}] & & time cut [\si{\ns}] & \\
  \cmidrule{3-5}
  \multirow{2}{*}{Charged particles} & & $0.75 \leq \pT{} < 4.00 $ & &  $t < 10.0$ & \\
  & & $0.00 \leq \pT{} < 0.75 $ & &  $t < 5.00$ & \\
  \bottomrule
\end{tabular}


\centering
\caption{\textit{Selected} selections used for the \SI{380}{\GeV} CLIC stage.}
\label{tab:LE_sel}

\begin{tabular}{lccccc}
  \toprule
  &
  \multicolumn{2}{c}{$|\cos\theta| \leq 0.975$} & &
  \multicolumn{2}{c}{$|\cos\theta| > 0.975$} \\ \cmidrule{2-3} \cmidrule{5-6}
  &
  \pT{} range [\si{\GeV}]& time cut [\si{\ns}] & &
  \pT{} range [\si{\GeV}]& time cut [\si{\ns}] \\
  \cmidrule{2-3}  \cmidrule{5-6}
  \multirow{2}{*}{Photons} &
  $0.75 \leq \pT{} < 2.00 $ & $t < 5.0$ &  &
  $0.75 \leq \pT{} < 2.00 $ & $t < 2.0$ \\
   &
  $0.00 \leq \pT{} < 0.75 $ & $t < 1.0$ & &
  $0.00 \leq \pT{} < 0.75 $ & $t < 1.0$ \\
  \cmidrule{2-3}  \cmidrule{5-6}
  \multirow{2}{*}{Neutral hadrons} &
  $0.75 \leq \pT{} < 2.00 $ & $t < 5.0$ &  &
  $0.75 \leq \pT{} < 2.00 $ & $t < 4.0$ \\
   &
  $0.00 \leq \pT{} < 0.75 $ & $t < 2.5$ & &
  $0.00 \leq \pT{} < 0.75 $ & $t < 2.0$ \\
  \cmidrule{2-6}
  &
  \multicolumn{5}{c}{$0 \leq |\cos\theta| \leq 1$} \\  \cmidrule{3-5}
  & &  \pT{} range [\si{\GeV}] & & time cut [\si{\ns}] & \\
  \cmidrule{3-5}
  \multirow{2}{*}{Charged particles} & & $0.75 \leq \pT{} < 4.00 $ & &  $t < 10.0$ & \\
  & & $0.00 \leq \pT{} < 0.75 $ & &  $t < 3.00$ & \\
  \bottomrule
\end{tabular}


\centering
\caption{\textit{Tight} selections used for the \SI{380}{\GeV} CLIC stage.}
\label{tab:LE_tight}

\begin{tabular}{lccccc}
  \toprule
  &
  \multicolumn{2}{c}{$|\cos\theta| \leq 0.975$} & &
  \multicolumn{2}{c}{$|\cos\theta| > 0.975$} \\ \cmidrule{2-3} \cmidrule{5-6}
  &
  \pT{} range [\si{\GeV}]& time cut [\si{\ns}] & &
  \pT{} range [\si{\GeV}]& time cut [\si{\ns}] \\
  \cmidrule{2-3}  \cmidrule{5-6}
  \multirow{2}{*}{Photons} &
  $0.75 \leq \pT{} < 2.00 $ & $t < 1.0$ &  &
  $0.75 \leq \pT{} < 2.00 $ & $t < 2.0$ \\
   &
  $0.00 \leq \pT{} < 0.75 $ & $t < 1.0$ & &
  $0.00 \leq \pT{} < 0.75 $ & $t < 1.0$ \\
  \cmidrule{2-3}  \cmidrule{5-6}
  \multirow{2}{*}{Neutral hadrons} &
  $0.75 \leq \pT{} < 2.00 $ & $t < 4.0$ &  &
  $0.75 \leq \pT{} < 2.00 $ & $t < 2.0$ \\
   &
  $0.00 \leq \pT{} < 0.75 $ & $t < 2.0$ & &
  $0.00 \leq \pT{} < 0.75 $ & $t < 2.0$ \\
  \cmidrule{2-6}
  &
  \multicolumn{5}{c}{$0 \leq |\cos\theta| \leq 1$} \\  \cmidrule{3-5}
  & &  \pT{} range [\si{\GeV}] & & time cut [\si{\ns}] & \\
  \cmidrule{3-5}
  \multirow{2}{*}{Charged particles} & & $0.75 \leq \pT{} < 3.00 $ & &  $t < 4.0$ & \\
  & & $0.00 \leq \pT{} < 0.75 $ & &  $t < 2.00$ & \\
  \bottomrule
\end{tabular}

\end{table}

\begin{figure}
  \centering
  \begin{subfigure}[b]{0.48\textwidth}
    \includegraphics[width=\textwidth]{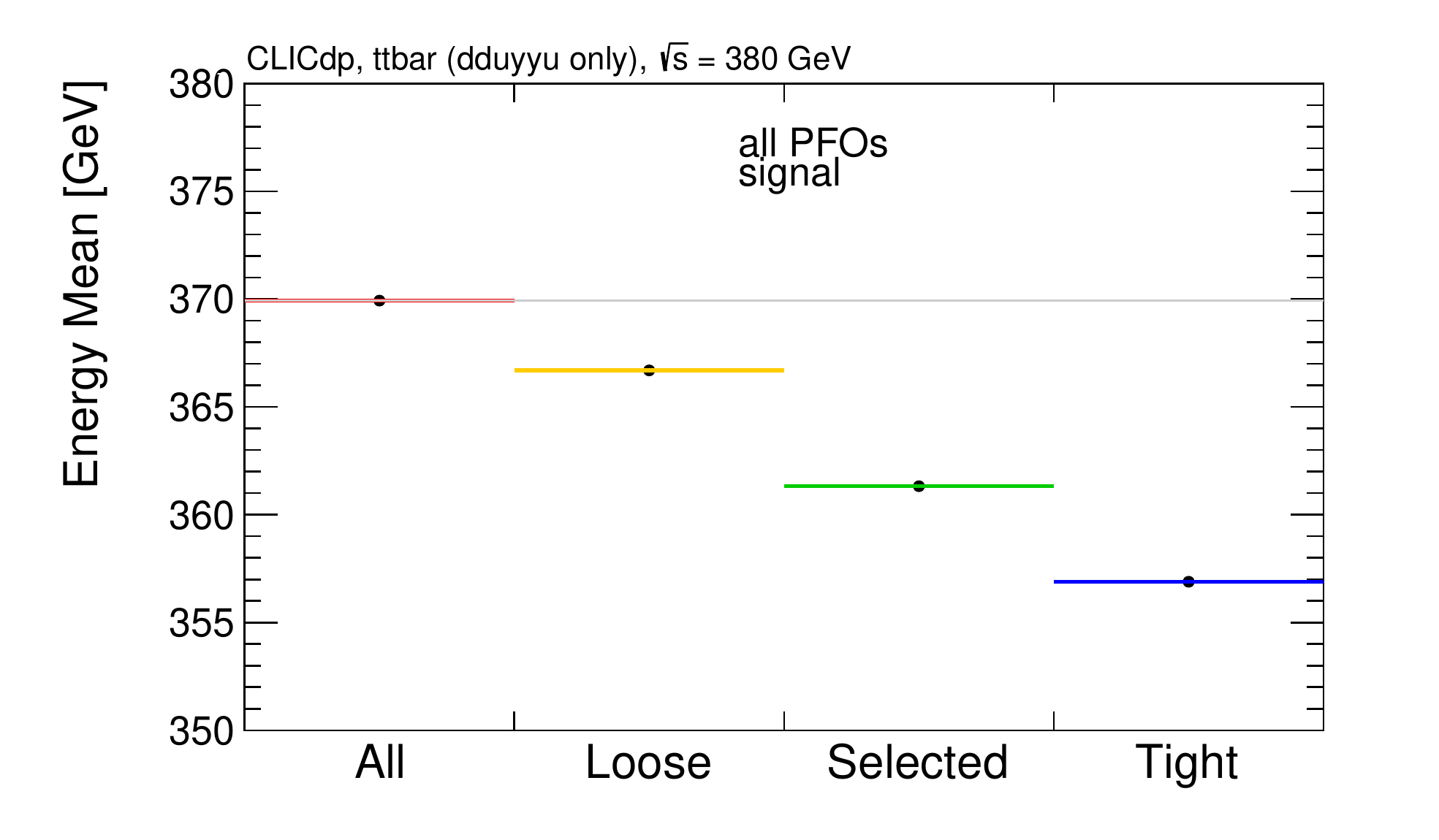}
  \end{subfigure}
  \hfill
  \begin{subfigure}[b]{0.48\textwidth}
    \includegraphics[width=\textwidth]{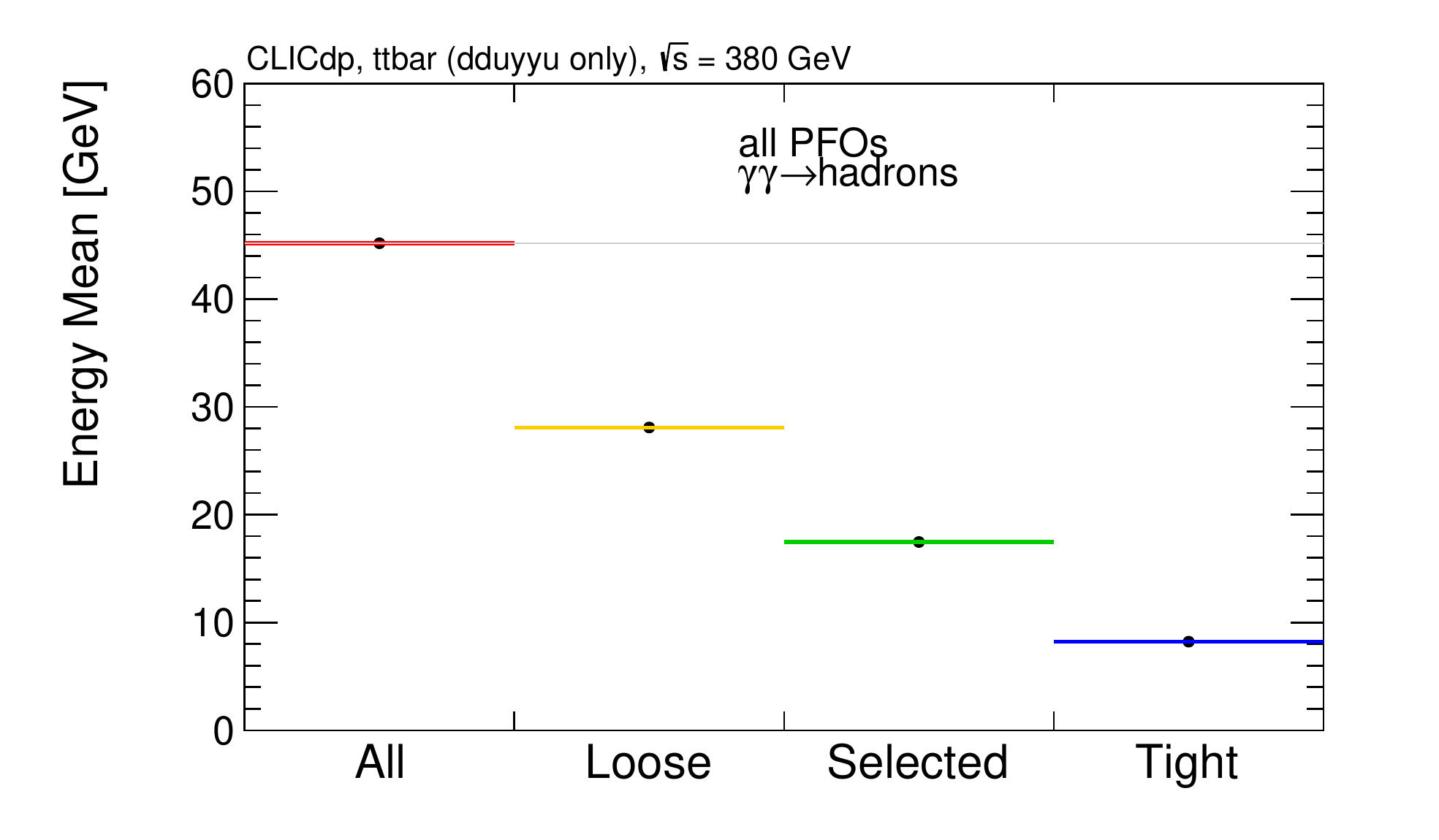}
  \end{subfigure}
  \caption{Signal and background reconstructed energy mean for different applied timing selections: without any selection (red),
  \textit{Loose} (orange), \textit{Selected} (green) and \textit{Tight} (blue). 
  \ttbar events are used with overlay of 30 BX of \gghads{} background for the \SI{380}{\GeV} CLIC stage.}
  \label{fig:energyMean_ttbar_380GeV_sig_ove}
\end{figure}

\begin{figure}
  \centering
  \begin{subfigure}[b]{0.48\textwidth}
    \includegraphics[width=\textwidth]{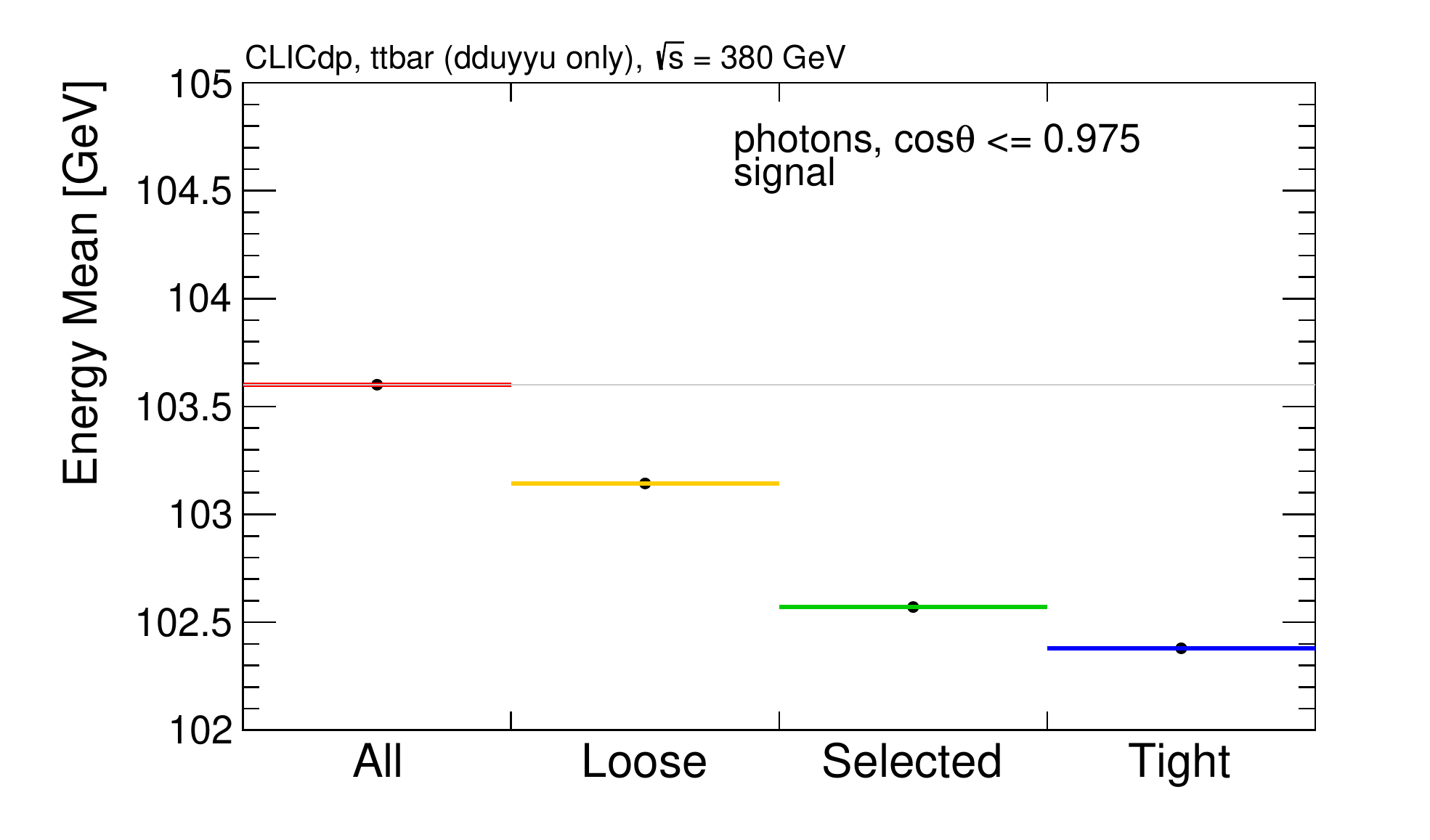}
  \end{subfigure}
  \hfill
  \begin{subfigure}[b]{0.48\textwidth}
    \includegraphics[width=\textwidth]{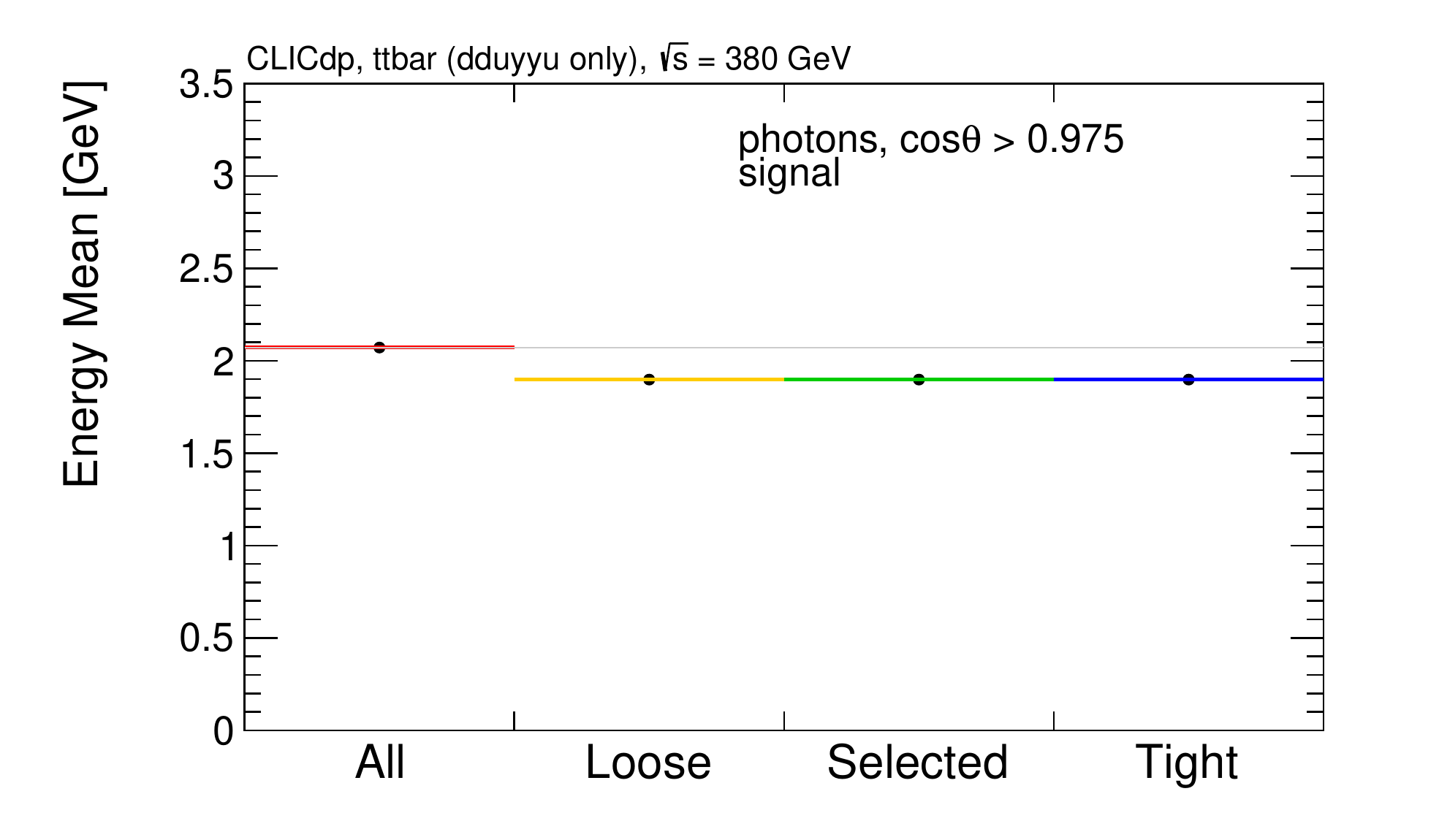}
  \end{subfigure}
  \begin{subfigure}[b]{0.48\textwidth}
    \includegraphics[width=\textwidth]{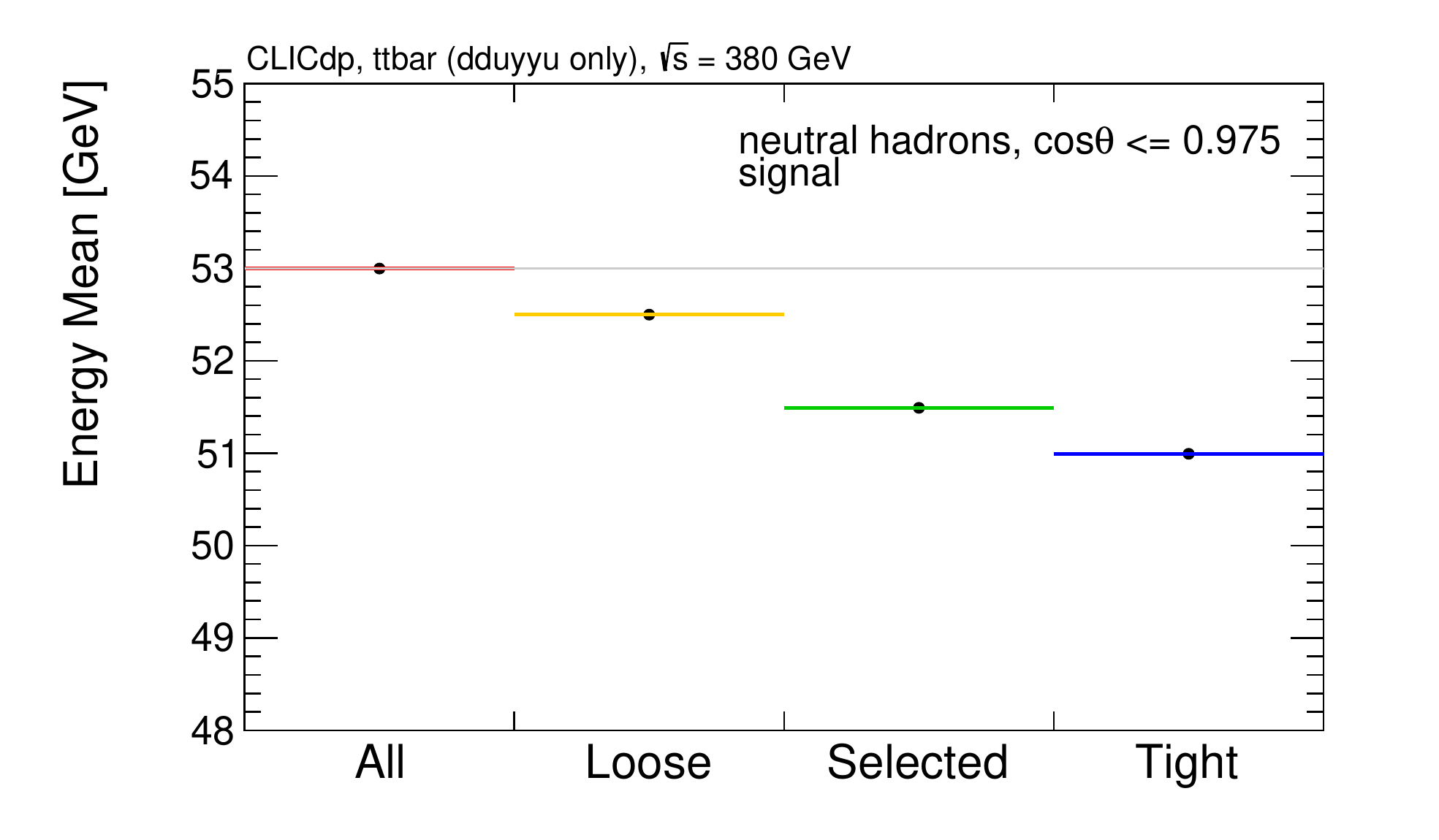}
  \end{subfigure}
  \hfill
  \begin{subfigure}[b]{0.48\textwidth}
    \includegraphics[width=\textwidth]{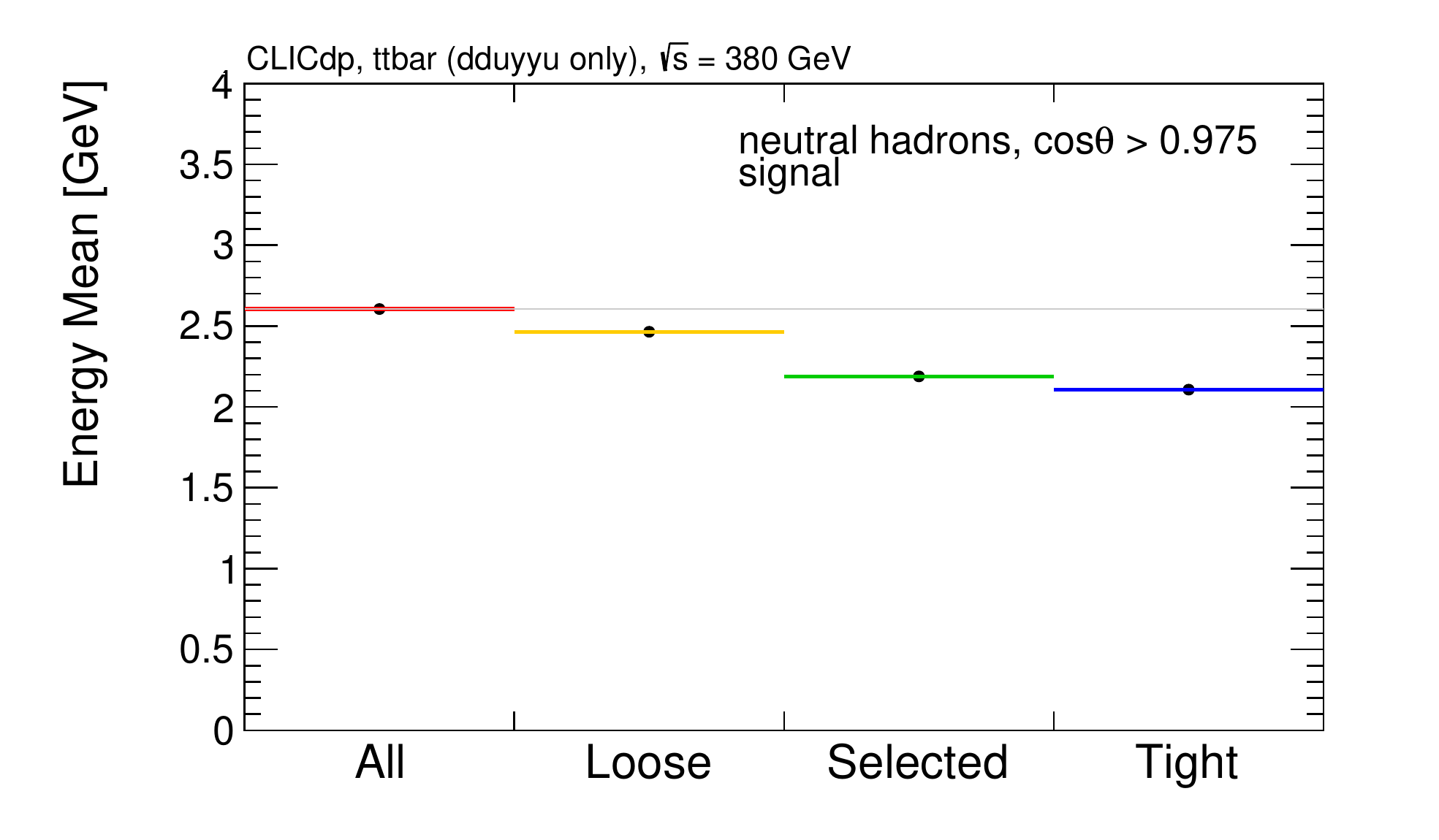}
  \end{subfigure}
  \begin{subfigure}[b]{0.48\textwidth}
    \includegraphics[width=\textwidth]{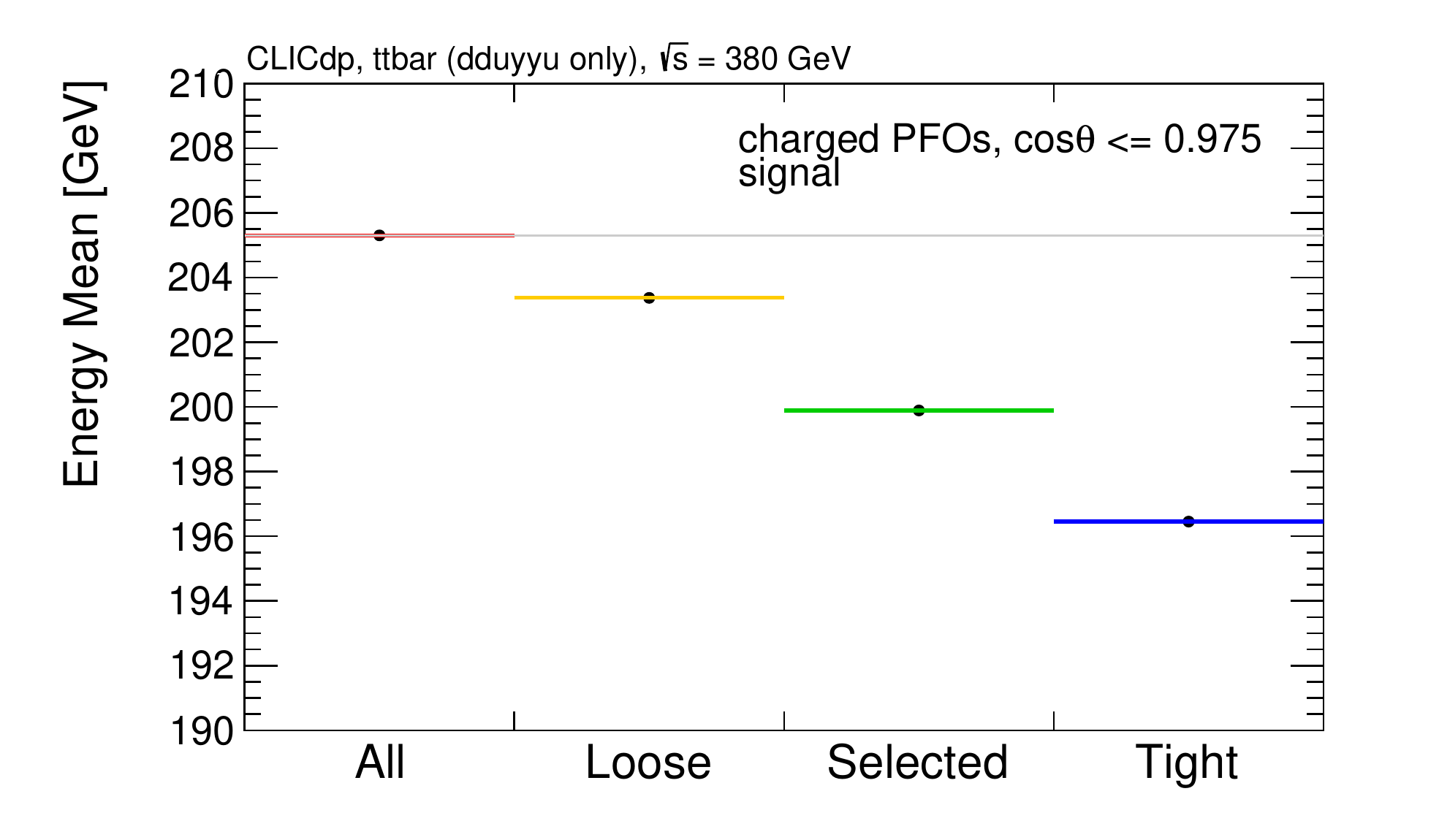}
  \end{subfigure}
  \hfill
  \begin{subfigure}[b]{0.48\textwidth}
    \includegraphics[width=\textwidth]{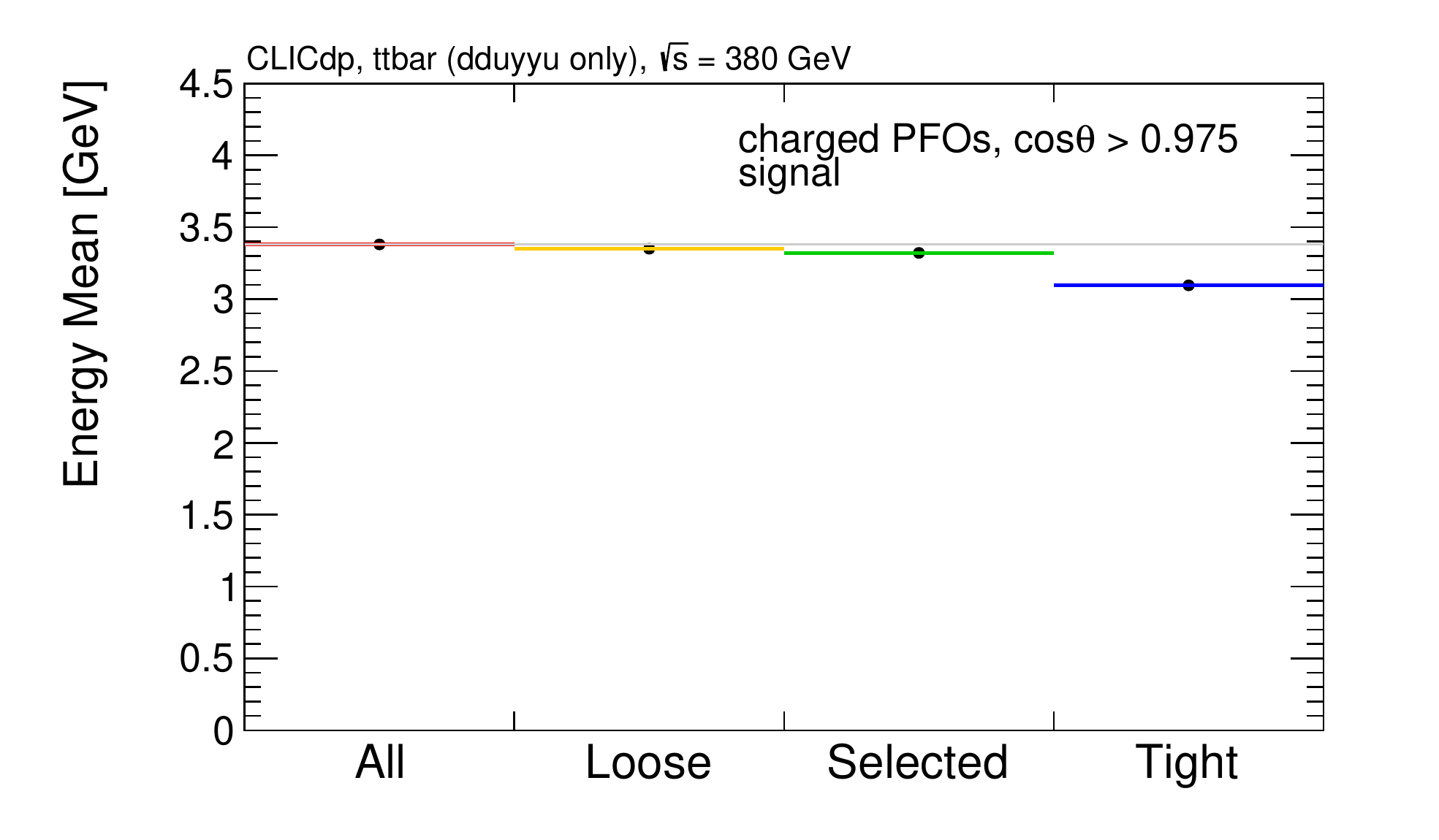}
  \end{subfigure}
  \caption{Signal reconstructed energy mean for the three particle categories (photons at the top,
  neutral hadrons in the middle and charged PFOs at the bottom) for different applied timing selections: without any selection (red),
  \textit{Loose} (orange), \textit{Selected} (green) and \textit{Tight} (blue). 
  \ttbar events are used with overlay of 30 BX of \gghads{} background for the \SI{380}{\GeV} CLIC stage.}
  \label{fig:energyMean_ttbar_380GeV_signal}
\end{figure}

\begin{figure}
  \centering
  \begin{subfigure}[b]{0.48\textwidth}
    \includegraphics[width=\textwidth]{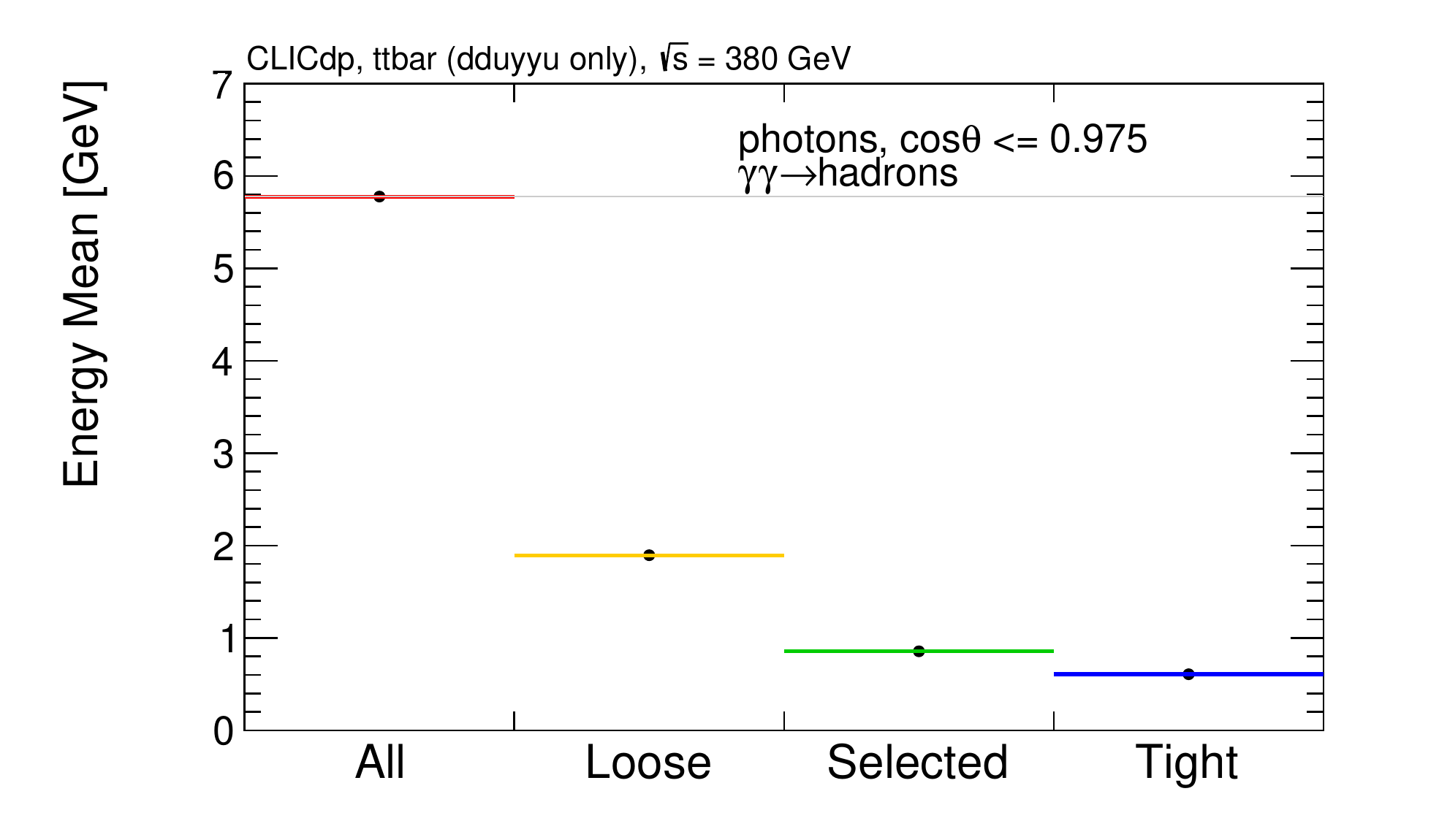}
  \end{subfigure}
  \hfill
  \begin{subfigure}[b]{0.48\textwidth}
    \includegraphics[width=\textwidth]{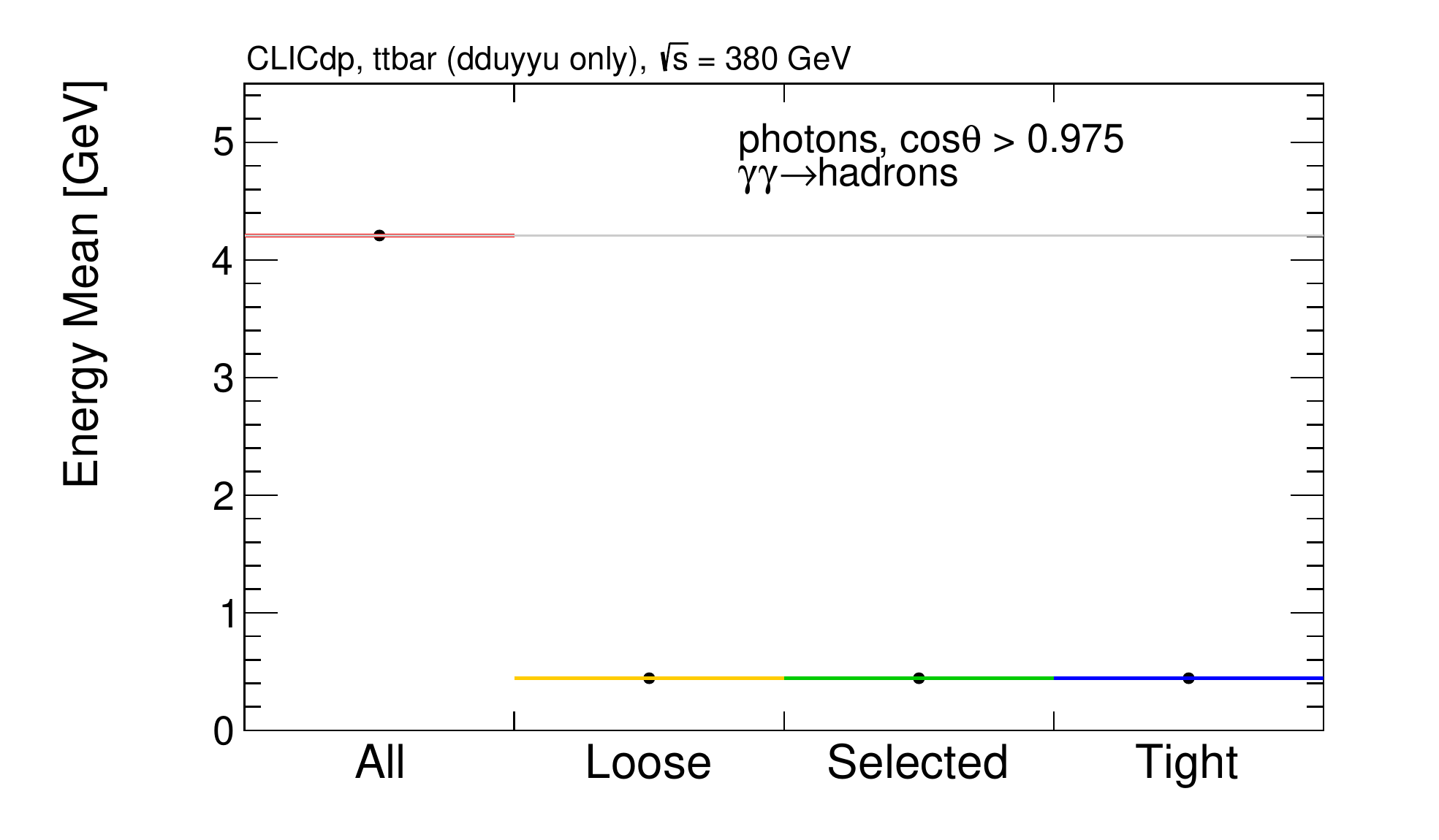}
  \end{subfigure}
  \begin{subfigure}[b]{0.48\textwidth}
    \includegraphics[width=\textwidth]{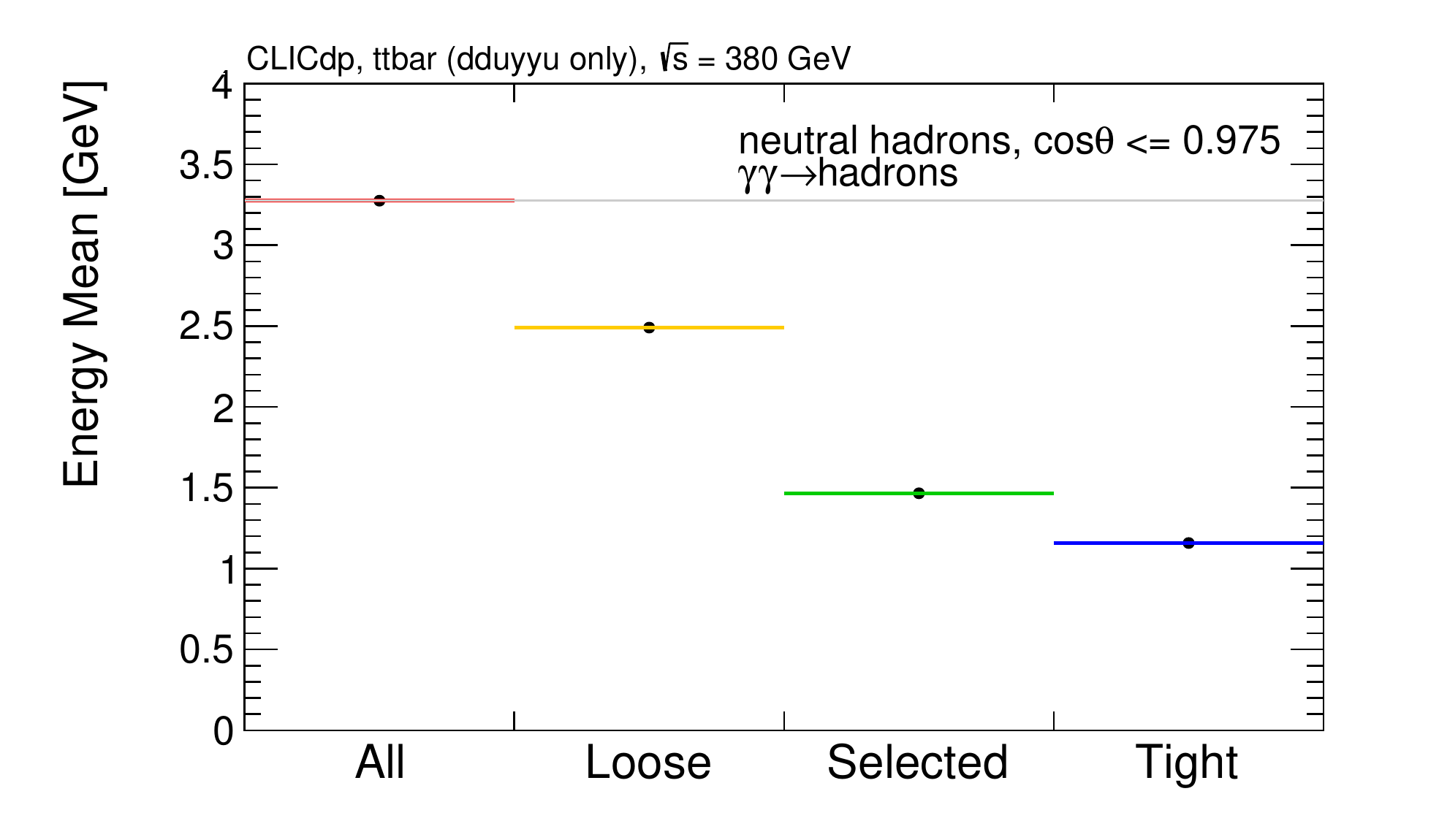}
  \end{subfigure}
  \hfill
  \begin{subfigure}[b]{0.48\textwidth}
    \includegraphics[width=\textwidth]{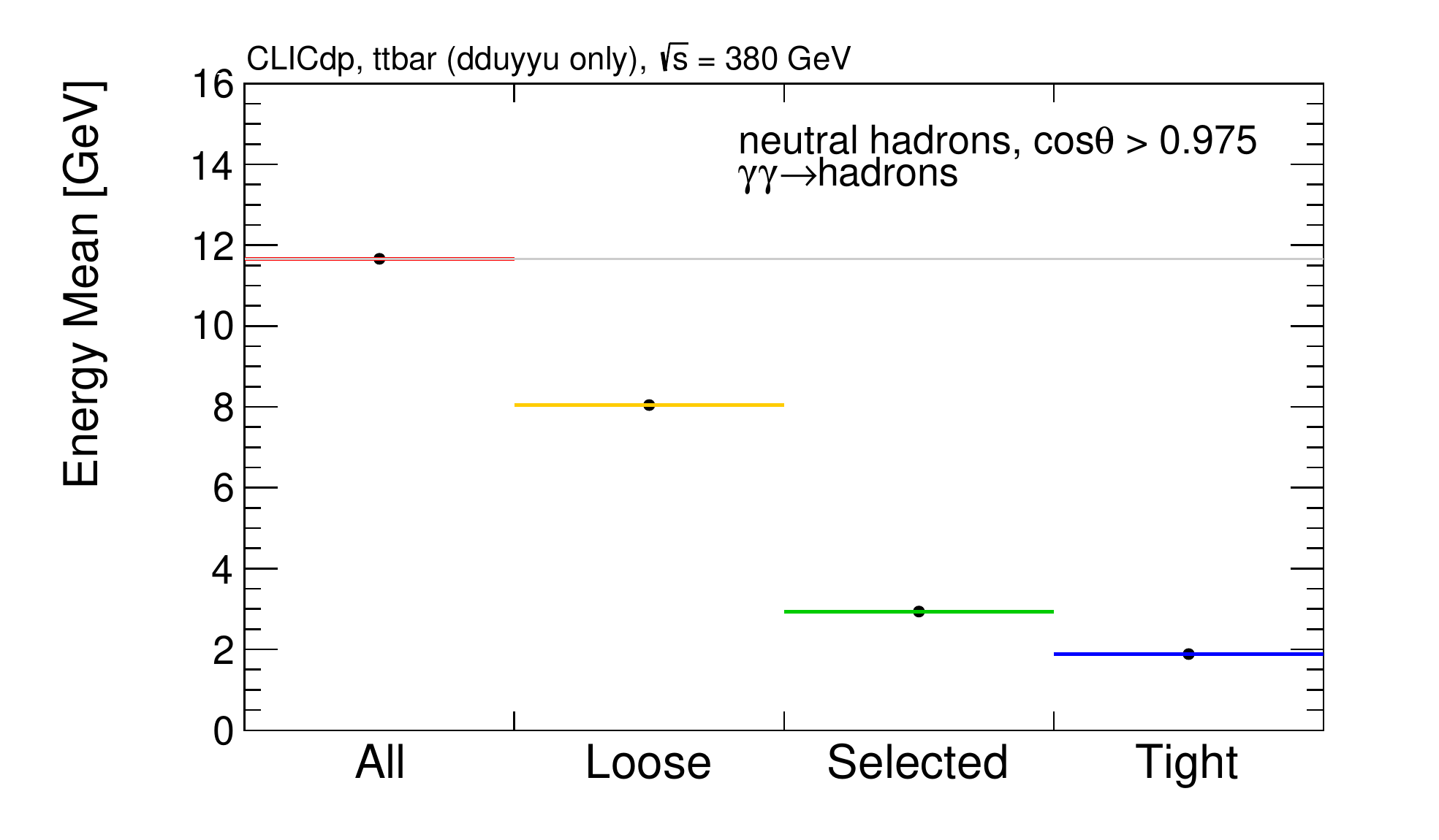}
  \end{subfigure}
  \begin{subfigure}[b]{0.48\textwidth}
    \includegraphics[width=\textwidth]{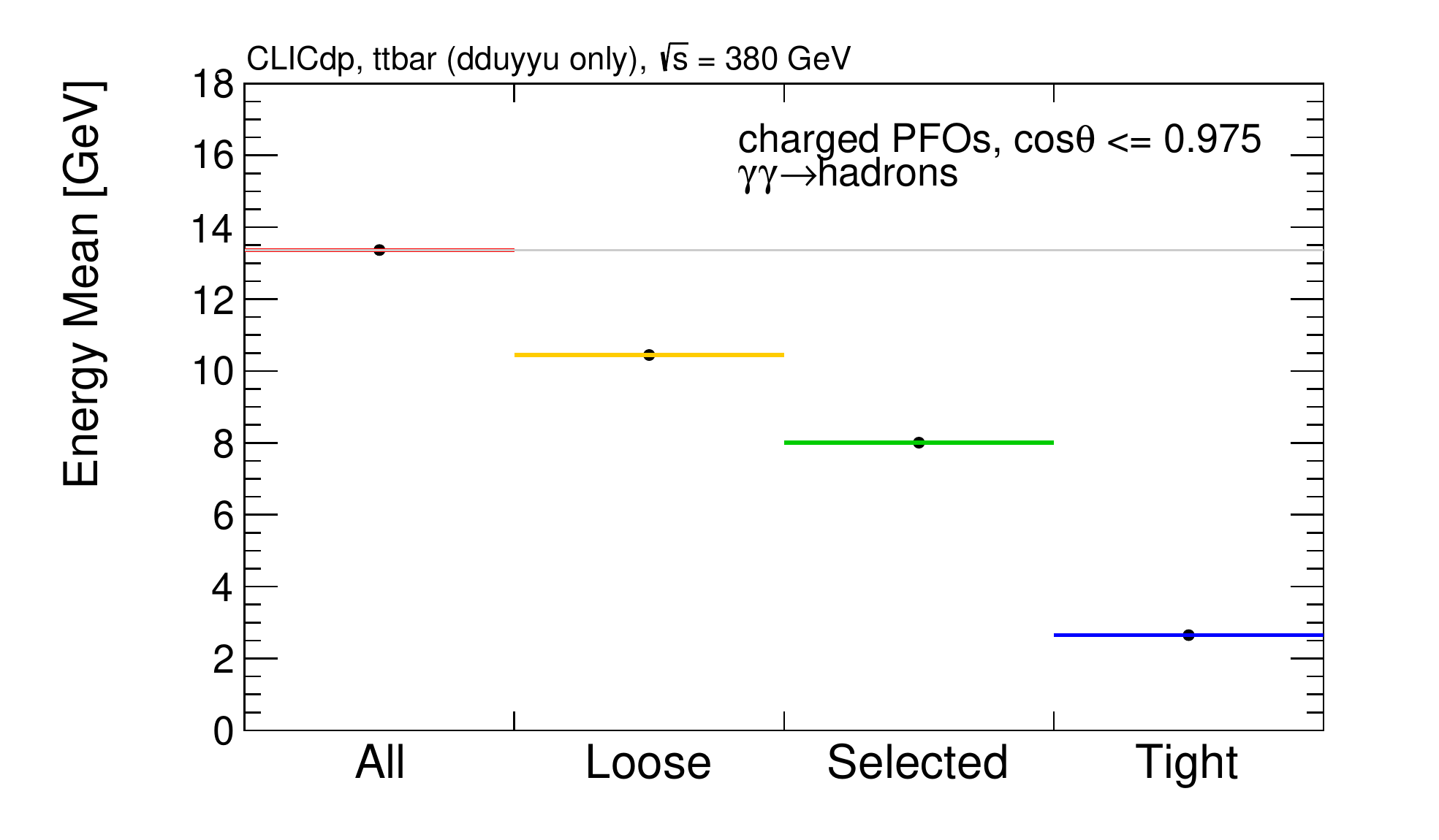}
  \end{subfigure}
  \hfill
  \begin{subfigure}[b]{0.48\textwidth}
    \includegraphics[width=\textwidth]{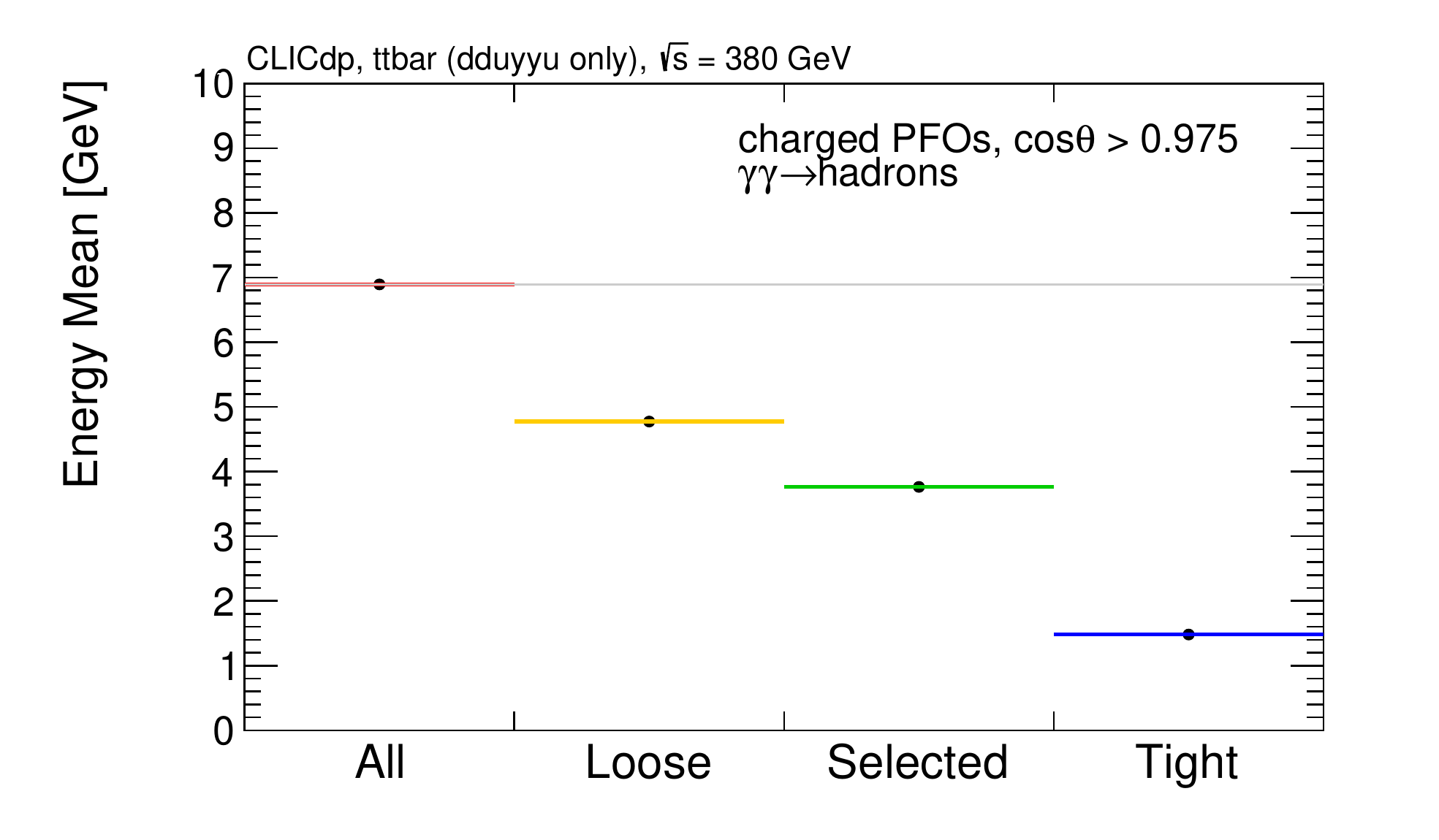}
  \end{subfigure}
  \caption{Background reconstructed energy mean for the three particle categories (photons at the top,
  neutral hadrons in the middle and charged PFOs at the bottom) for different applied timing selections: without any selection (red),
  \textit{Loose} (orange), \textit{Selected} (green) and \textit{Tight} (blue). 
  \ttbar events are used with overlay of 30 BX of \gghads{} background for the \SI{380}{\GeV} CLIC stage.}
  \label{fig:energyMean_ttbar_380GeV_ove}
\end{figure}

The mean of the energy distribution for the overall distribution of signal and background 
is shown in~\autoref{fig:energyMean_ttbar_380GeV_sig_ove}. 
The mean of the energy distribution after the selections is shown 
in~\cref{fig:energyMean_ttbar_380GeV_signal,fig:energyMean_ttbar_380GeV_ove}
for each particle category and polar angle group
for the signal component and the overlay, respectively. 
It can be noted that for PFOs with $\cos\theta > 0.975$ the reduction of the background is obtained with almost no loss in
the signal, in particular in the case of photons.
The level of background is reduced more with the \textit{Tight} selection than with the other two selections, but more signal is lost. 
In the case of the charged PFOs with $\cos\theta \leq 0.975$,
the signal loss is almost \SI{10}{\GeV} in the extreme case of the \textit{Tight} selection.
When considering the \textit{Loose} selection, the mean of the signal energy distribution is shifted less than \SI{1}{\GeV} 
in the case of photons and neutral hadrons, while in the case of charged PFOs it is almost $\SI{2}{\GeV}$.
Therefore, in the next section, the \pT{} vs. time selections on the charged PFOs are modified in the attempt of recovering this loss.

\section{Optimization of timing selections for the \SI{380}{\GeV} CLIC stage}
\label{sec:LEtimingCuts}

To check for possible optimizations, the following relaxed \textit{Loose} selections are applied:
\begin{itemize}
\item increasing the time cut from \SI{5}{\ns} to \SI{6}{\ns} in the charged PFO category, referred to as \textit{time$+$};
\item increasing the time cut from \SI{5}{\ns} to \SI{8}{\ns} in the charged PFO category, referred to as \textit{time$++$};
\item decreasing the \pT{} cut from \SI{0.75}{\GeV} to \SI{0.5}{\GeV} for all categories, referred to as \textit{\pT{}$-$};
\item decreasing the \pT{} cut from \SI{0.75}{\GeV} to \SI{0}{\GeV} for all categories, referred to as \textit{\pT{}$--$}.
\end{itemize}
They are compared to the case where no selection is applied and to the \textit{Loose} selections listed in~\autoref{tab:LE_loose}, 
also referred to as \textit{vanilla} selections.

\begin{figure}
  \centering
  \begin{subfigure}[b]{0.48\textwidth}
    \includegraphics[width=\textwidth]{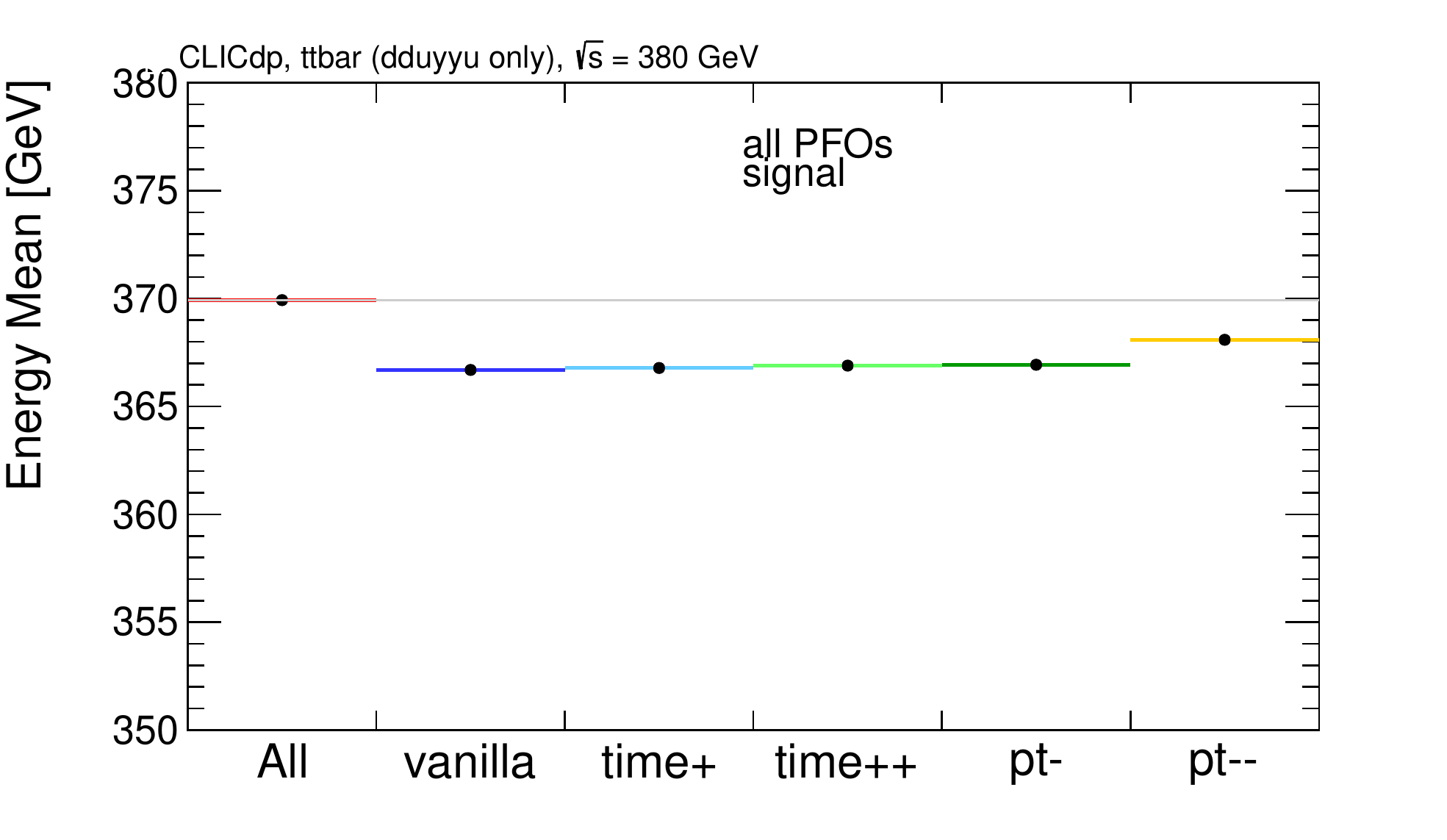}
  \end{subfigure}
  \hfill
  \begin{subfigure}[b]{0.48\textwidth}
    \includegraphics[width=\textwidth]{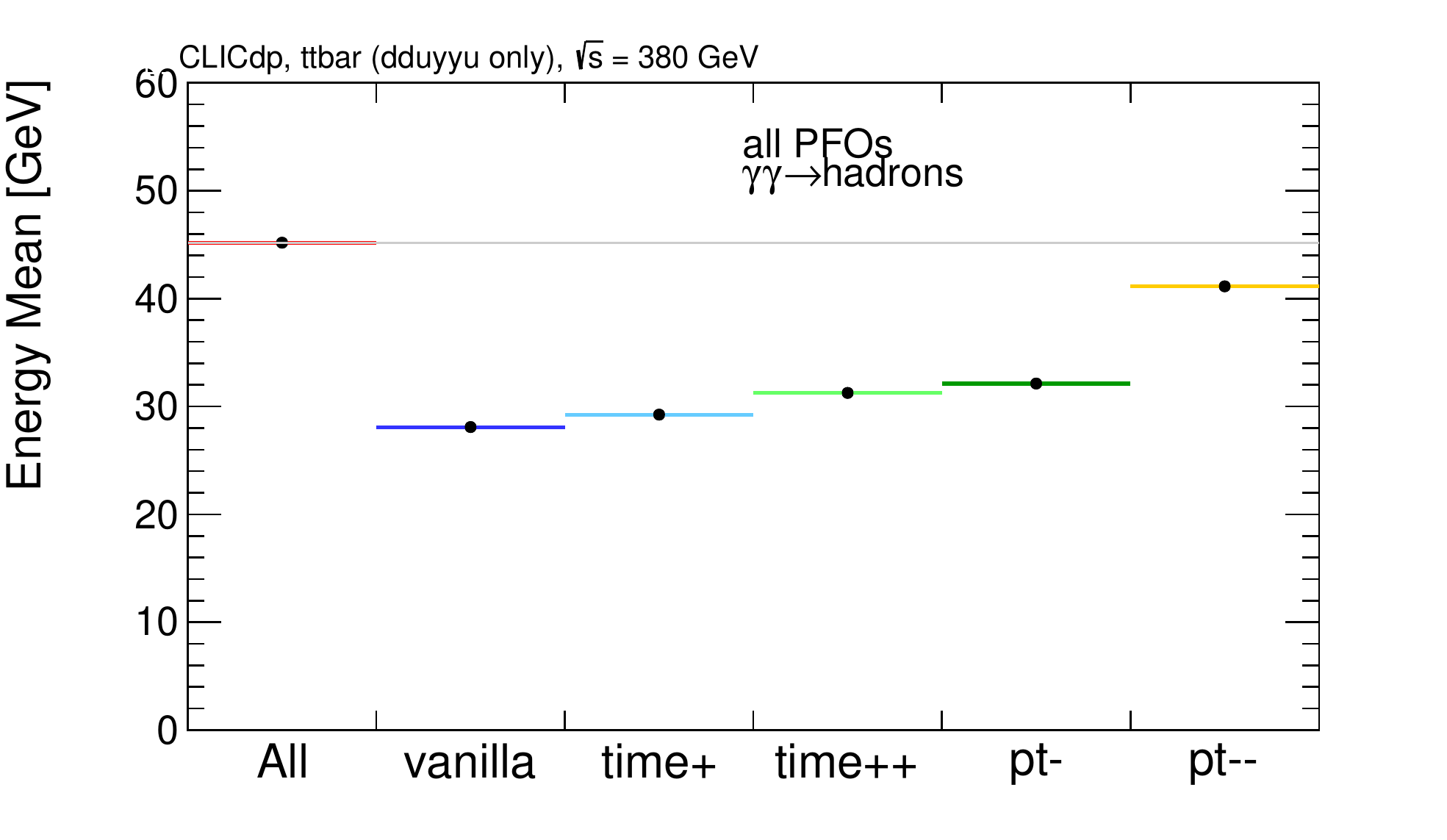}
  \end{subfigure}
  \caption{Signal and background reconstructed energy mean for different relaxed cuts on the \textit{Loose} selection: time$+$,
  time$++$, \pT{}$-$, \pT{}$--$, from left to right respectively.
  \ttbar events are used with overlay of 30 BX of \gghads{} background for the \SI{380}{\GeV} CLIC stage.}
  \label{fig:energyMean_ttbar_380GeV_diffSel}
\end{figure}

\begin{figure}
  \centering
  \begin{subfigure}[b]{0.48\textwidth}
    \includegraphics[width=\textwidth]{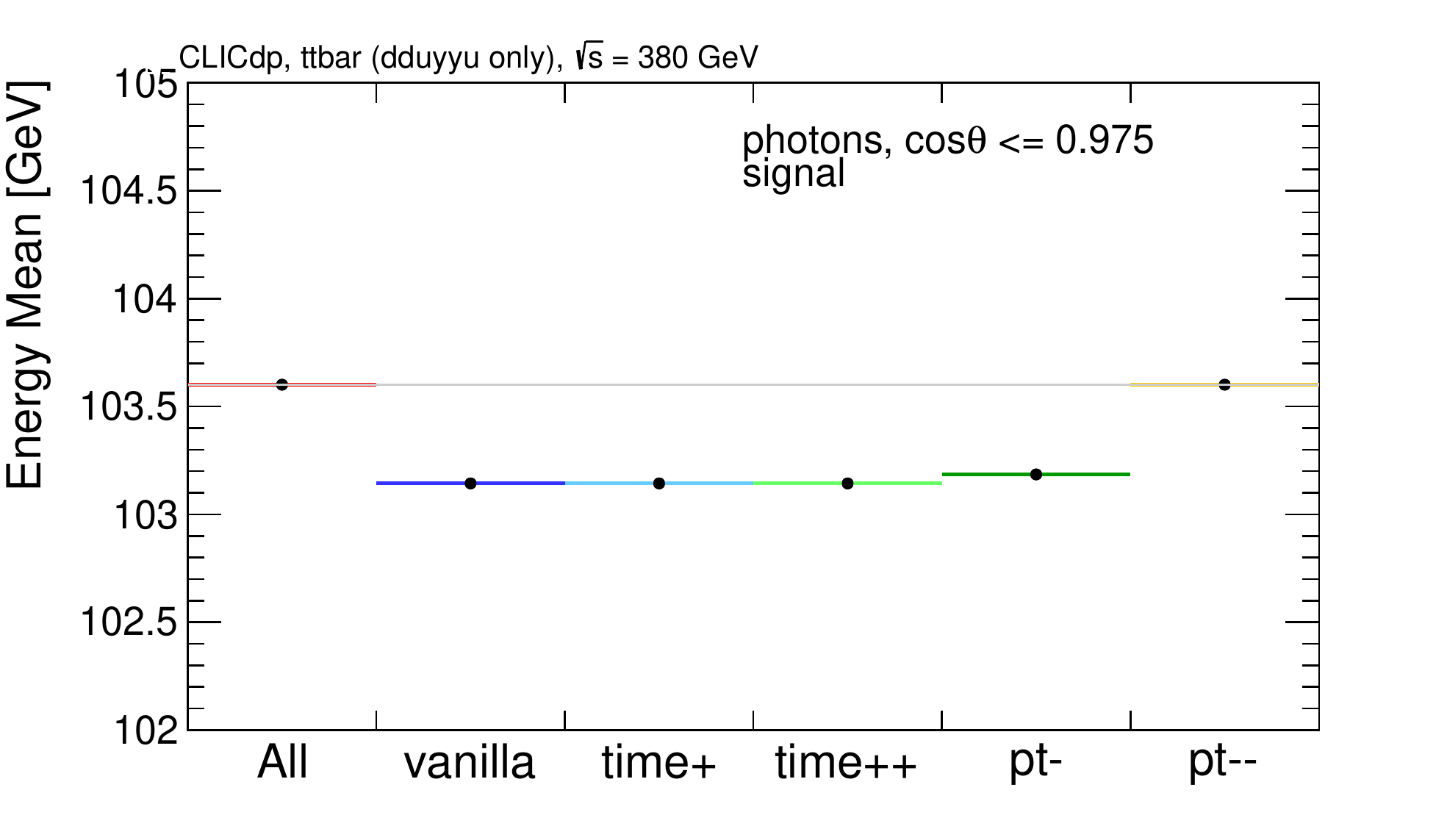}
  \end{subfigure}
  \hfill
  \begin{subfigure}[b]{0.48\textwidth}
    \includegraphics[width=\textwidth]{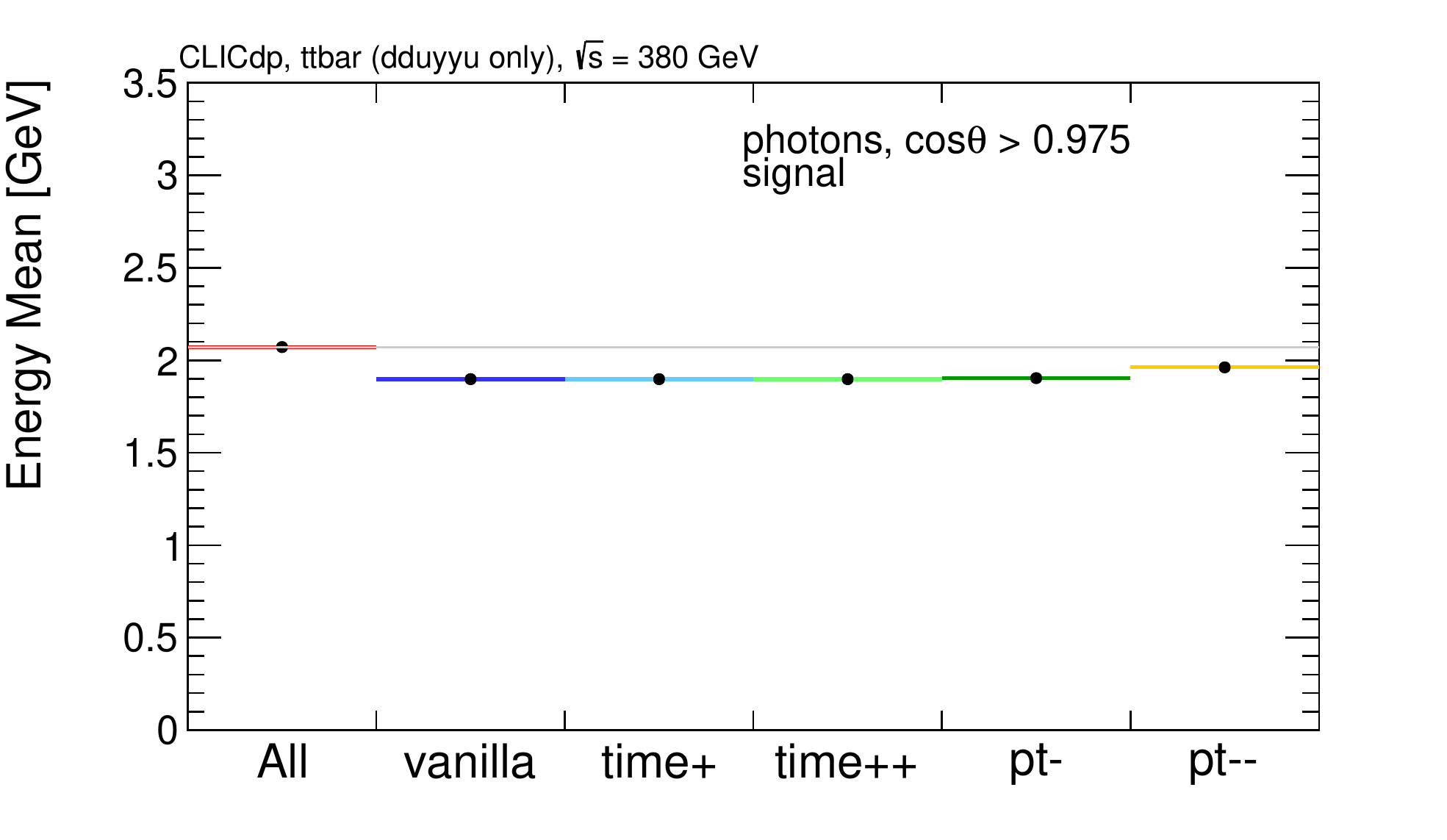}
  \end{subfigure}
  \begin{subfigure}[b]{0.48\textwidth}
    \includegraphics[width=\textwidth]{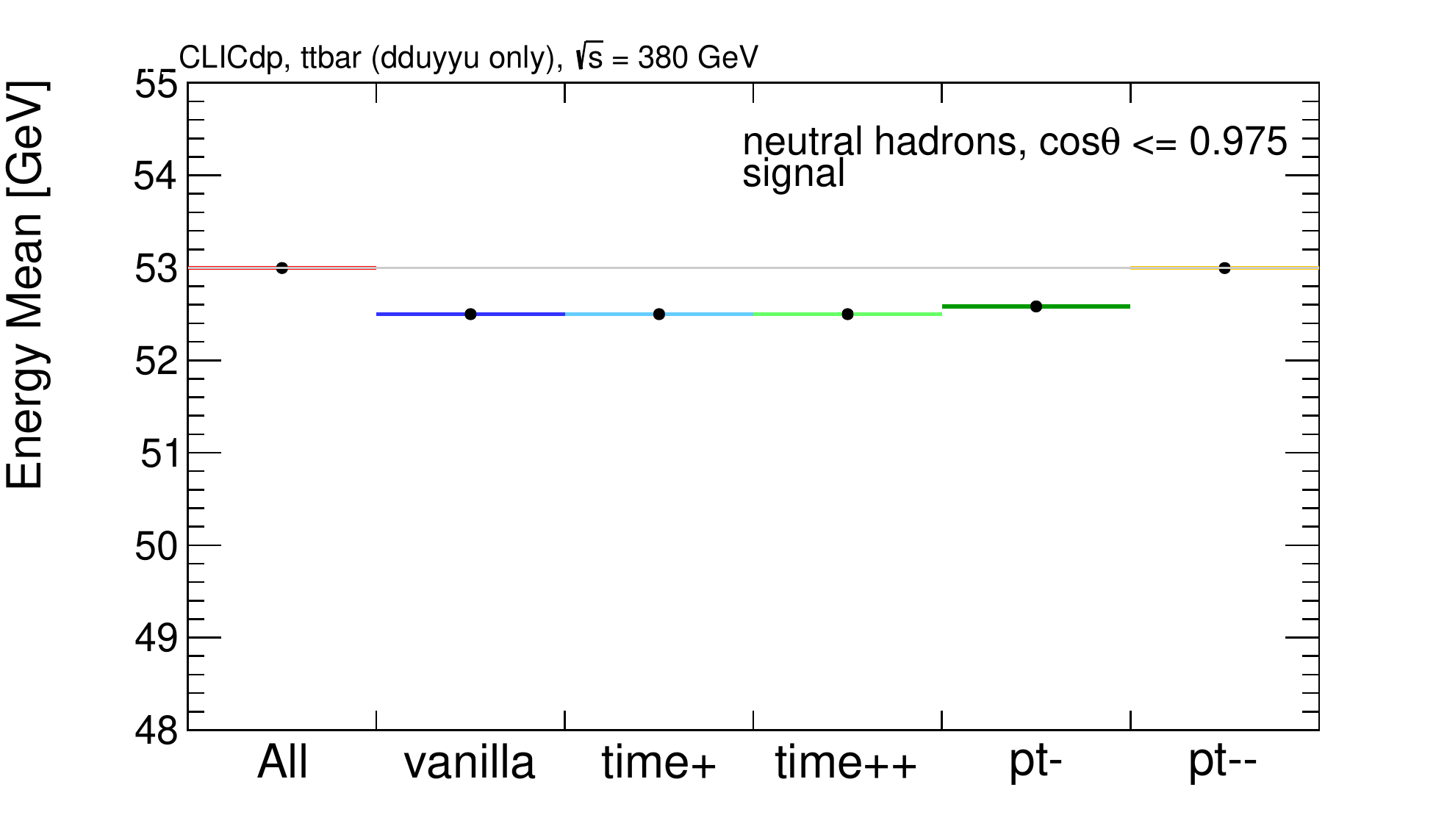}
  \end{subfigure}
  \hfill
  \begin{subfigure}[b]{0.48\textwidth}
    \includegraphics[width=\textwidth]{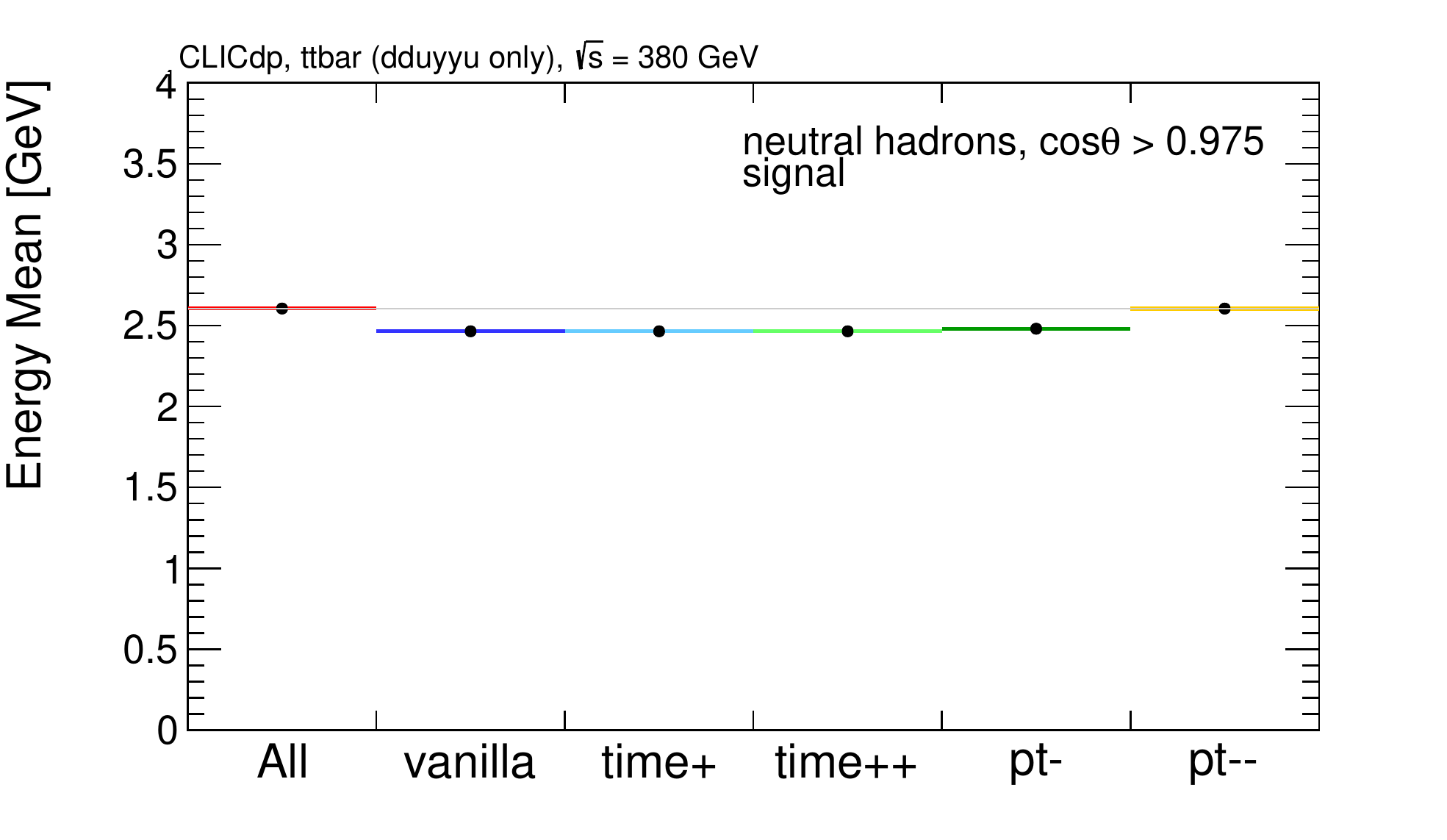}
  \end{subfigure}
  \begin{subfigure}[b]{0.48\textwidth}
    \includegraphics[width=\textwidth]{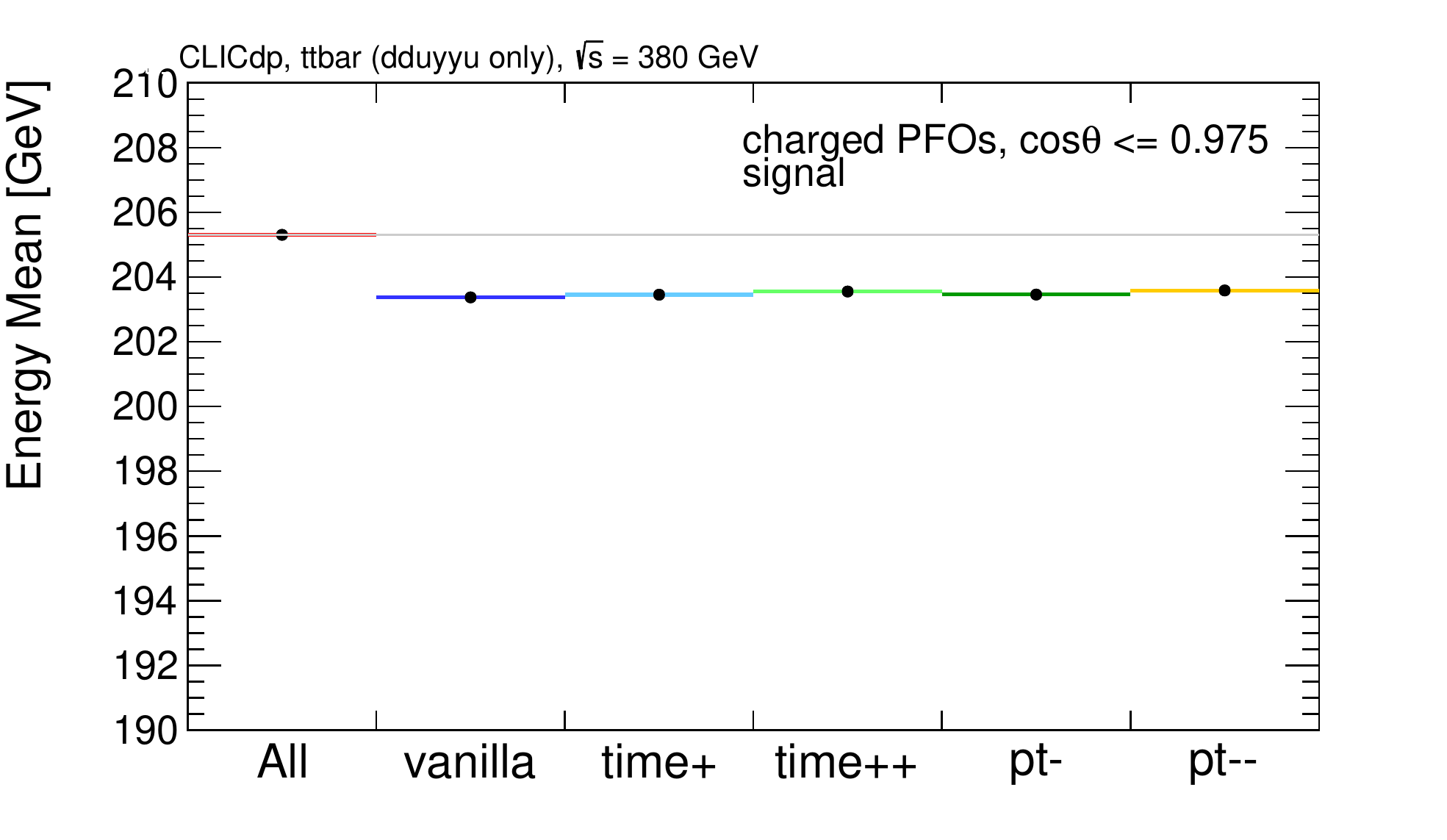}
  \end{subfigure}
  \hfill
  \begin{subfigure}[b]{0.48\textwidth}
    \includegraphics[width=\textwidth]{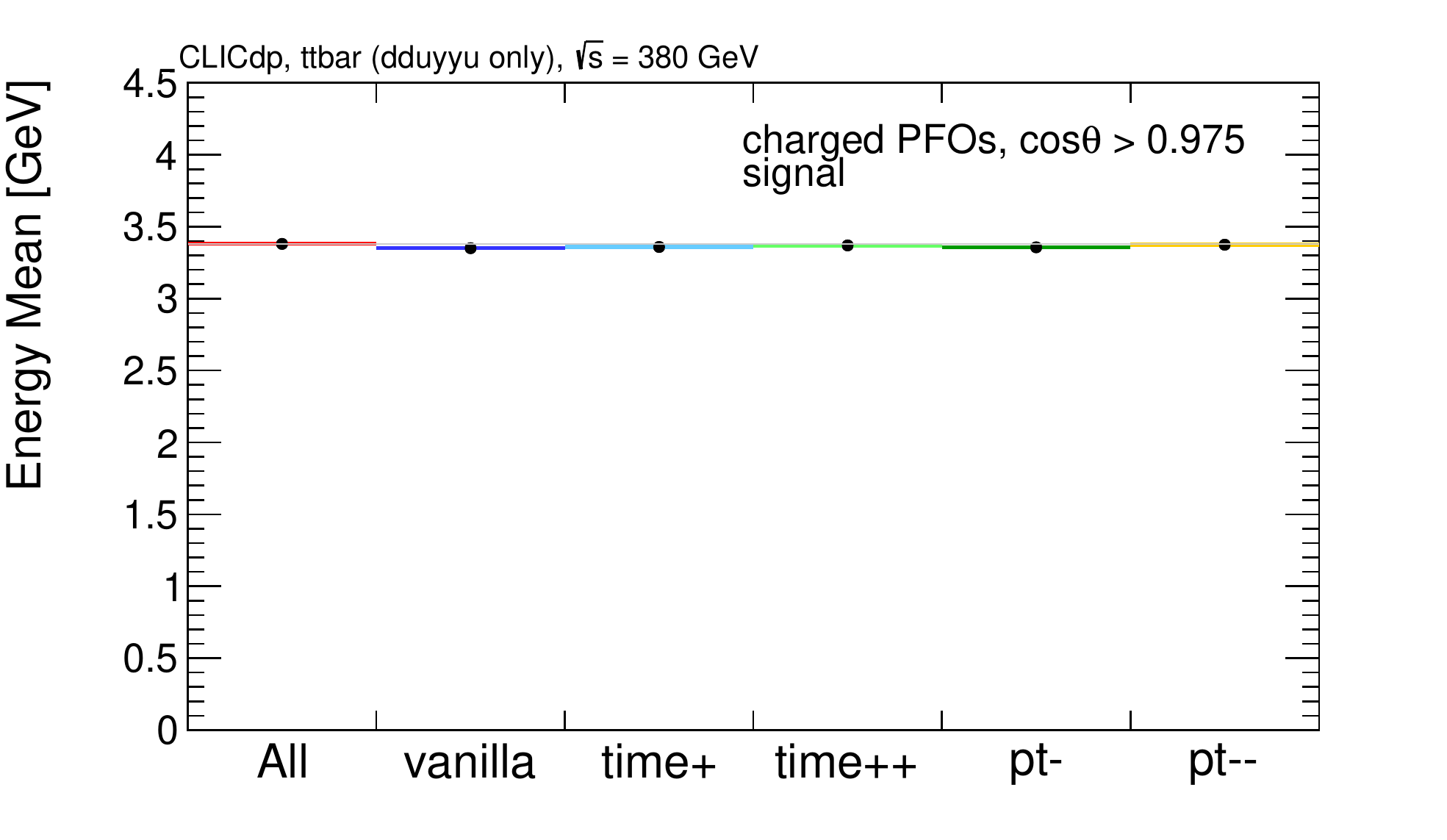}
  \end{subfigure}
  \caption{Signal reconstructed energy mean for the three particle categories (photons at the top,
  neutral hadrons in the middle and charged PFOs at the bottom) for different relaxed cuts on the \textit{Loose} selection: time$+$,
  time$++$, \pT{}$-$, \pT{}$--$, from left to right respectively.
  \ttbar events are used with overlay of 30 BX of \gghads{} background for the \SI{380}{\GeV} CLIC stage.}
  \label{fig:energyMean_ttbar_380GeV_signal_diffSel}
\end{figure}

\begin{figure}
  \centering
  \begin{subfigure}[b]{0.48\textwidth}
    \includegraphics[width=\textwidth]{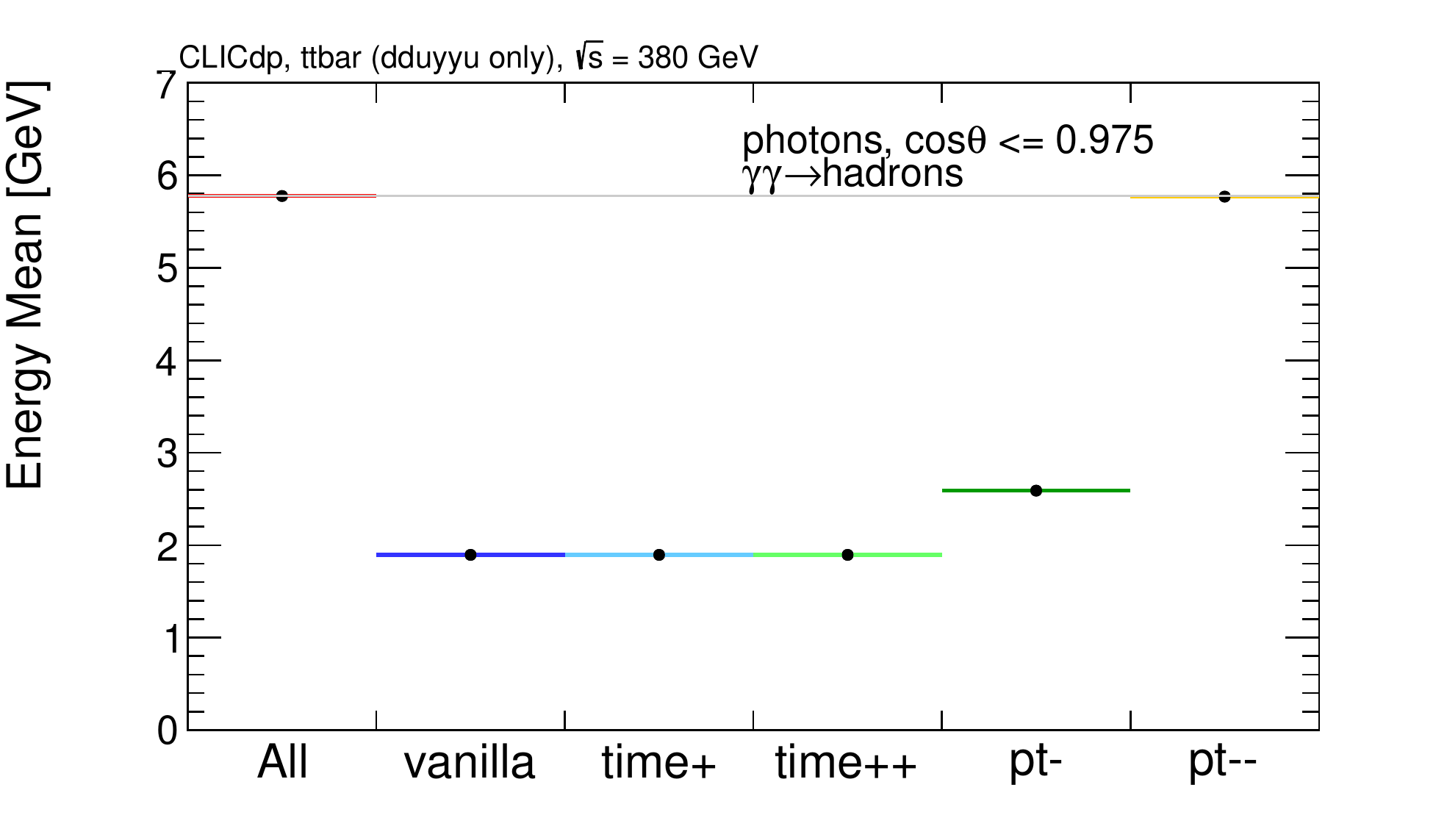}
  \end{subfigure}
  \hfill
  \begin{subfigure}[b]{0.48\textwidth}
    \includegraphics[width=\textwidth]{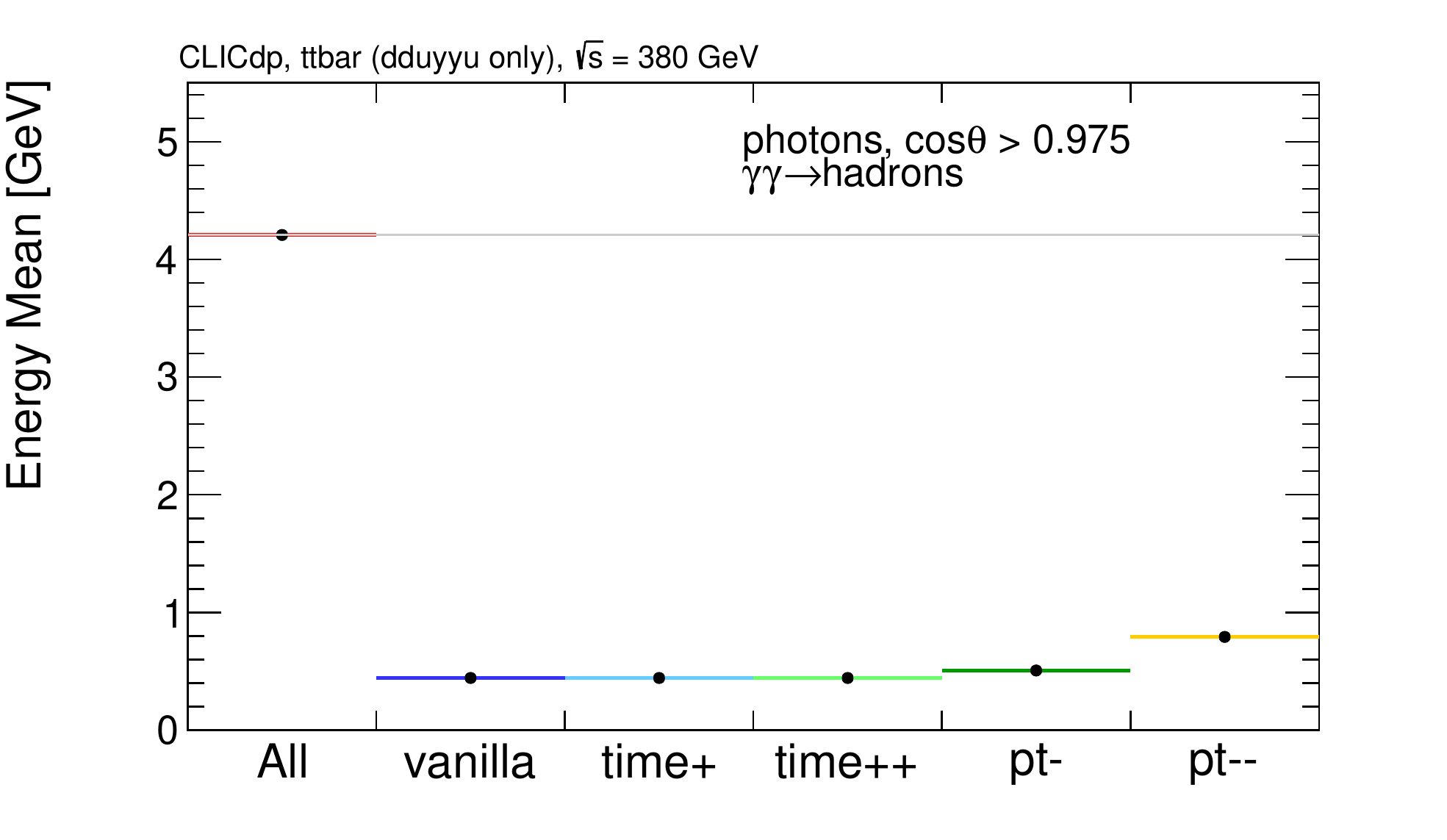}
  \end{subfigure}
  \begin{subfigure}[b]{0.48\textwidth}
    \includegraphics[width=\textwidth]{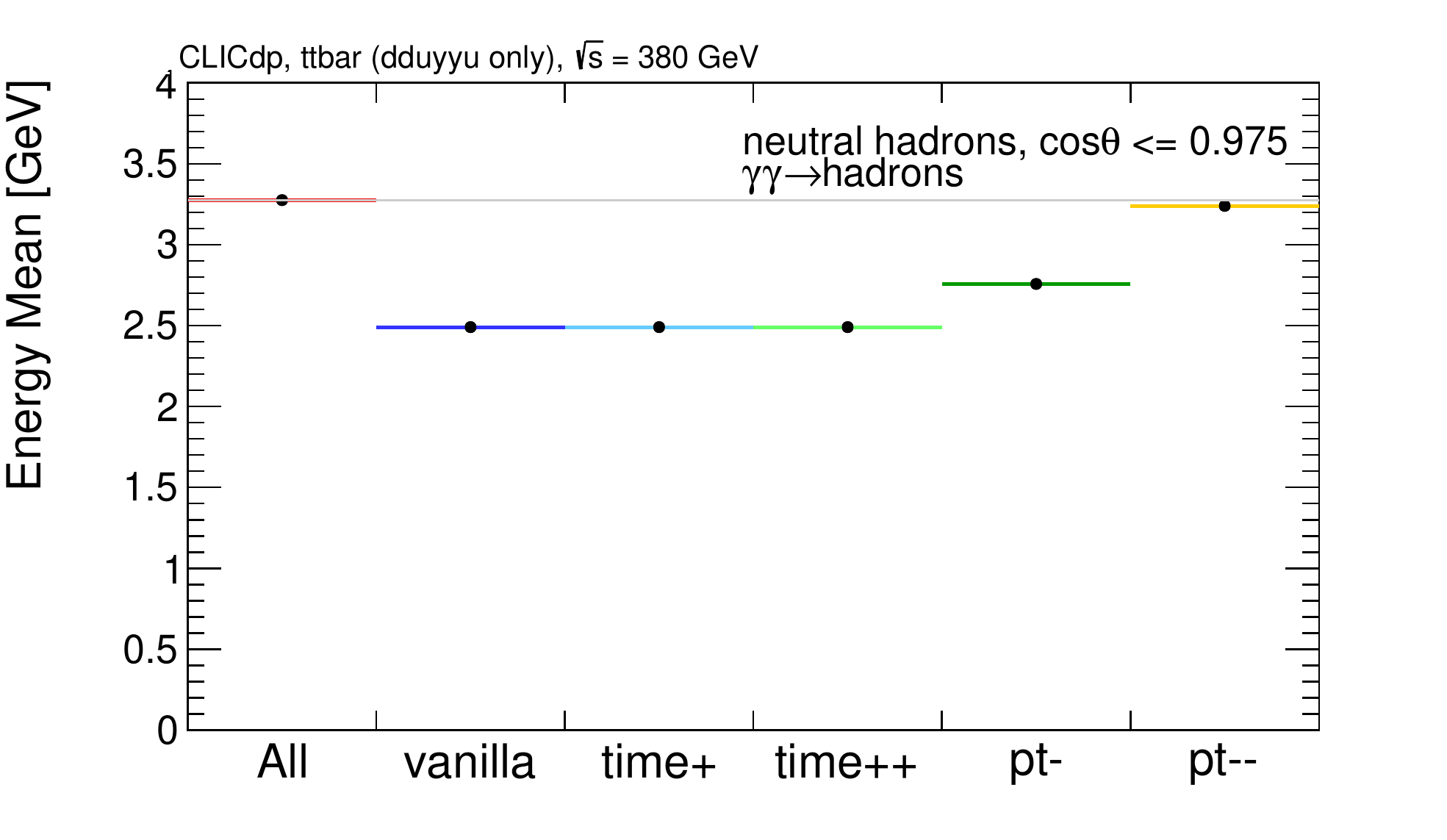}
  \end{subfigure}
  \hfill
  \begin{subfigure}[b]{0.48\textwidth}
    \includegraphics[width=\textwidth]{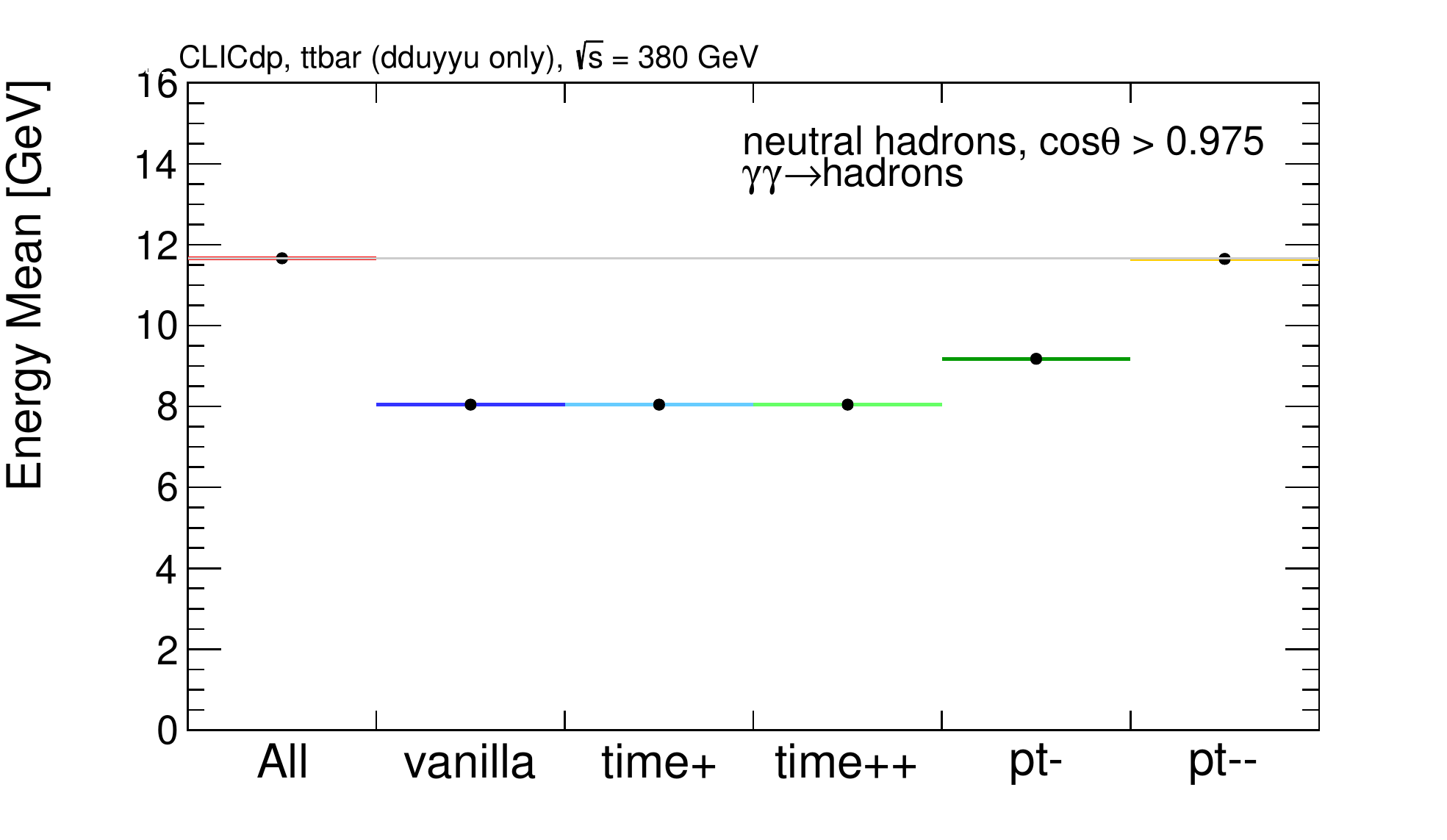}
  \end{subfigure}
  \begin{subfigure}[b]{0.48\textwidth}
    \includegraphics[width=\textwidth]{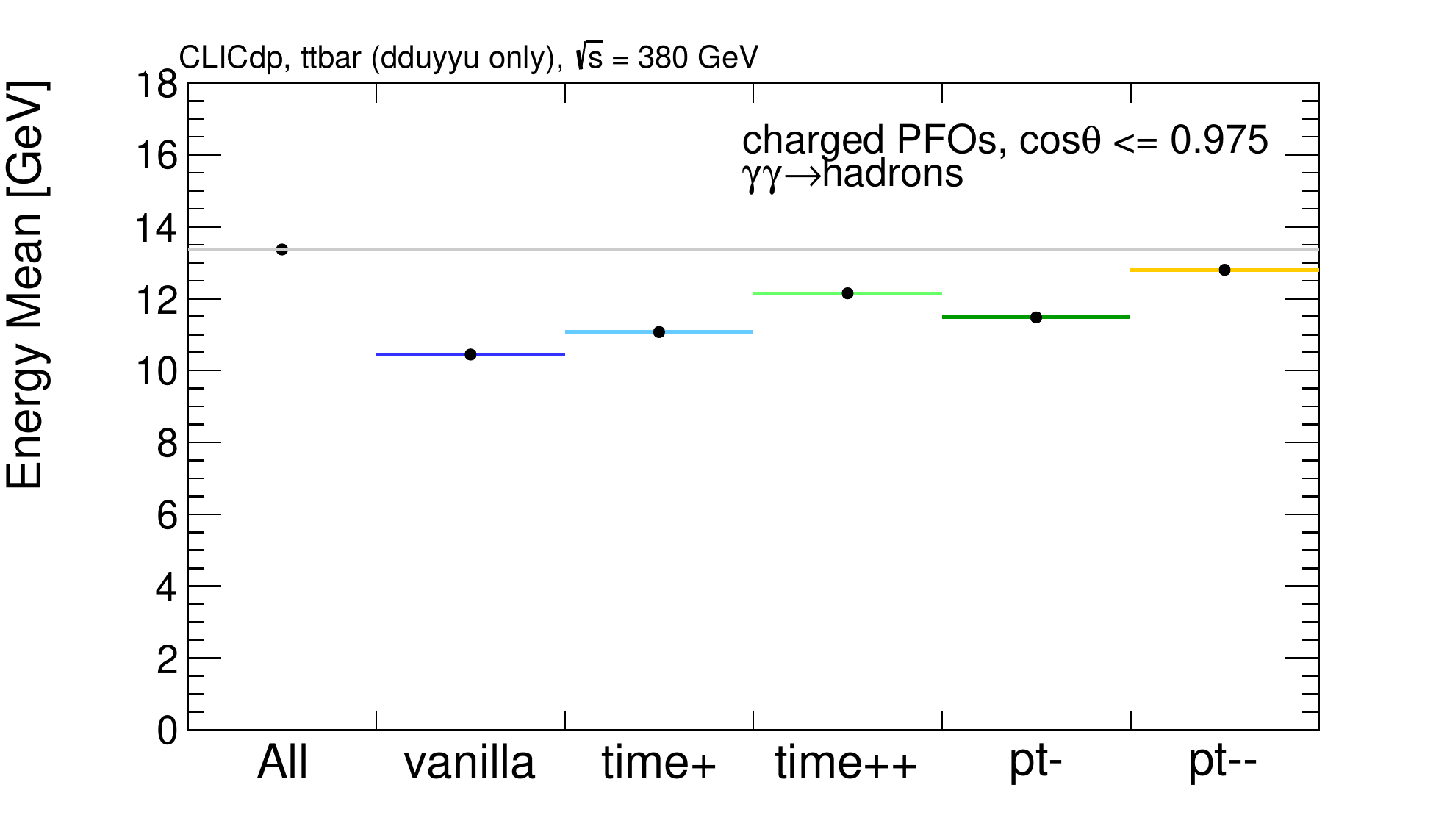}
  \end{subfigure}
  \hfill
  \begin{subfigure}[b]{0.48\textwidth}
    \includegraphics[width=\textwidth]{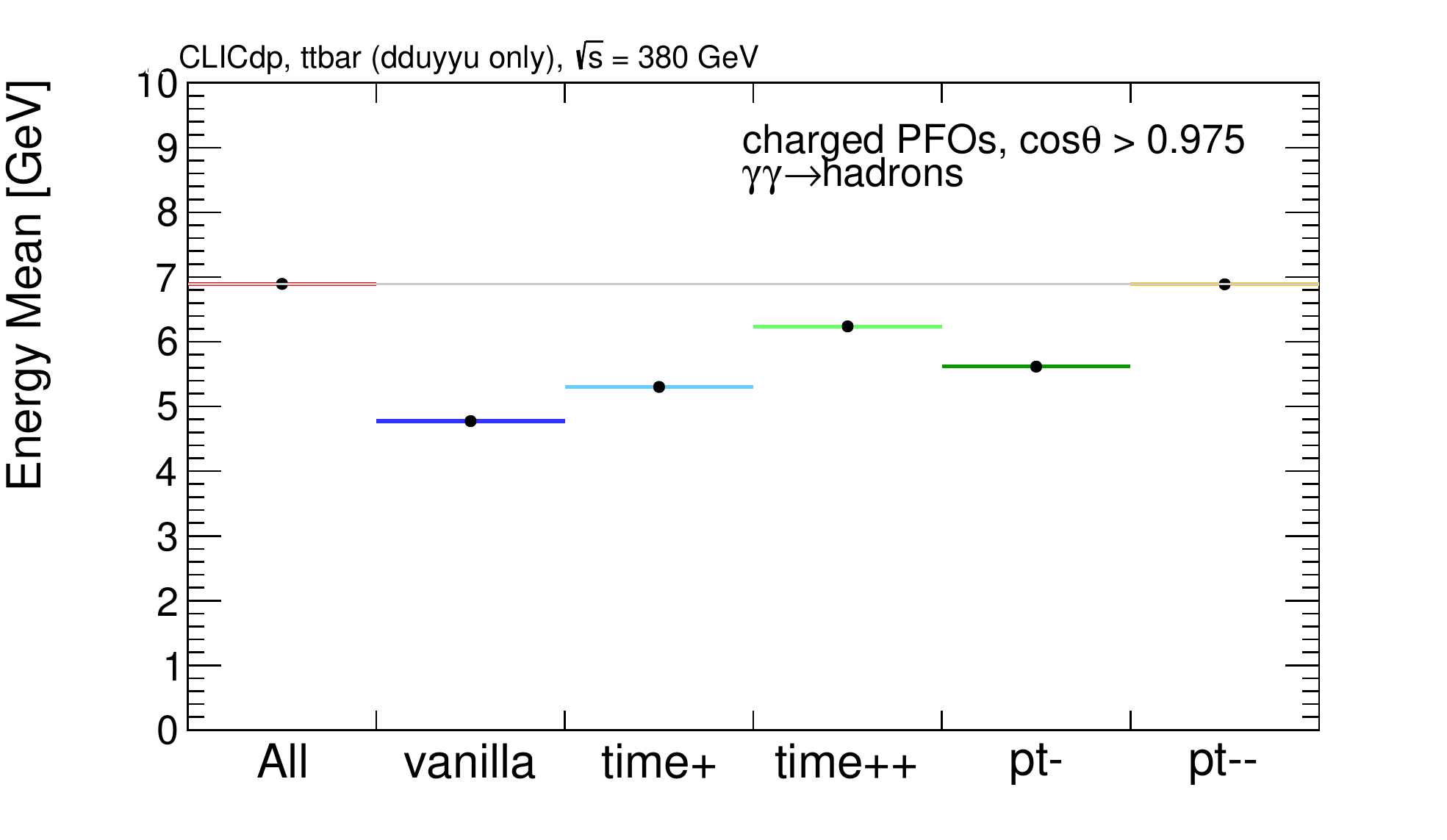}
  \end{subfigure}
  \caption{Background reconstructed energy mean for the three particle categories (photons at the top,
  neutral hadrons in the middle and charged PFOs at the bottom) for different relaxed cuts on the \textit{Loose} selection: time$+$,
  time$++$, \pT{}$-$, \pT{}$--$, from left to right respectively.
  \ttbar events are used with overlay of 30 BX of \gghads{} background for the \SI{380}{\GeV} CLIC stage.}
  \label{fig:energyMean_ttbar_380GeV_ove_diffSel}
\end{figure}

The comparison of the energy mean of the relaxed cuts to the \textit{vanilla} one can be found in~\autoref{fig:energyMean_ttbar_380GeV_diffSel}
for all PFOs contained in each \ttbar event and in~\cref{fig:energyMean_ttbar_380GeV_signal_diffSel,fig:energyMean_ttbar_380GeV_ove_diffSel} 
for the three particle categories divided into signal and background.
The effect of relaxing the cuts is an increase of the background component. 
The \pT{}$--$ selection is the only one recovering the signal component
but also leaving the level of background essentially unchanged with respect to not applying any cuts.
This last observation is valid only for photons and neutral hadrons, but not for charged PFOs.
In fact, part of the signal component of the charged PFOs suffers from a 
mis-correction of the PFO time in the case where the track and the calorimeter clusters associated with the charged PFO
are not produced by the same simulated particle. Therefore, this signal component cannot be recovered
with a simple modification to the \pT{} vs. time selections.
In conclusion, this study shows that no further optimization is possible on top of the \textit{Loose} selections currently
defined for the \SI{380}{\GeV} CLIC accelerator.

\section{Conclusions}
\label{sec:results}

In this study the performance of the timing selections defined for the CLIC \SI{380}{\GeV} energy stage is evaluated.
\ttbar events coming from hard \epem interactions and decaying mainly into light quarks are simulated 
and reconstructed as signal. At the same time a realistic estimate of the \gghads{} is included as background.
The mean of the total energy reconstructed in all events is chosen as a figure-of-merit for this study.
The results for the selection for the CLIC \SI{380}{\GeV} collider are found to be satisfactory:
after applying the selections, the level of background is significantly reduced down to a few~GeV, 
while the energy mean of the signal distribution is only few percent.
In the extreme case of the \textit{Tight} selection for example, the background is reduced from about \SI{45}{\GeV} down to about \SI{8}{\GeV}
and less than 4\% is lost in the signal energy mean.
In a second step of this study, the cuts in the \textit{Loose} selection are relaxed to search for possible improvements.
Among all the options considered, only the \pT{}$--$ relaxed selection partially recovers the signal component
but also leaves the background energy mean unchanged with respect to not applying any cuts.
Therefore, no change is foreseen on the timing selections for the CLIC \SI{380}{\GeV} energy stage.

\section*{Acknowledgements}
\label{sec:acknowledgements}

This work benefited from services provided by the ILC Virtual Organisation, supported by the national resource providers of the EGI Federation.
This research was done using resources provided by the Open Science Grid, which is supported by the National Science Foundation and the U.S. Department of Energy's Office of Science.

\newpage

\printbibliography[title=References]

\end{document}